\documentclass[12pt]{iopart}

\usepackage{cite}
\usepackage{graphicx}

\usepackage{color}
\usepackage{ulem}

\newcommand{\be}{\begin{eqnarray}}
\newcommand{\ee}{\end{eqnarray}}
\newcommand{\sub}[1]{_\mathrm{#1}}

\begin{document}

\title[Dynamic force spectroscopy of DNA hairpins. I.]{Dynamic force spectroscopy of DNA hairpins. I. Force kinetics and free energy landscapes}

\author{A~Mossa$^1$, M~Manosas$^1$\footnote{Present address:
    Laboratoire de Physique Statistique, Ecole Normale Sup\'erieure,
    Unit\'e Mixte de Recherche 8550 associ\'ee au Centre National de
    la recherche Scientifique et aux Universit\'es Paris VI et VII, 24
    Rue Lhomond, 75231 Paris, France.}, N~Forns$^{1,2}$,
  J~M~Huguet$^1$ and F~Ritort$^{1,2}$}

\address{$^1$ Departament de F\'{\i}sica Fonamental, Facultat de F\'{\i}sica, Universitat de Barcelona, Diagonal 647, 08028 Barcelona, Spain}
\address{$^2$ CIBER-BBN Networking center on Bioengineering, Biomaterials and Nanomedicine}
\ead{ritort@ffn.ub.es}

\begin{abstract}
We investigate the thermodynamics and kinetics of DNA hairpins that
fold/unfold under the action of applied mechanical force. We introduce
the concept of the molecular free energy landscape and derive
simplified expressions for the force dependent Kramers--Bell rates. To
test the theory we have designed a specific DNA hairpin sequence that
shows two-state cooperative folding under mechanical tension and
carried out pulling experiments using optical tweezers. We show
how we can determine the parameters that characterize the molecular
free energy landscape of such sequence from rupture force kinetic
studies. Finally we combine such kinetic studies with experimental
investigations of the Crooks fluctuation relation to derive the free
energy of formation of the hairpin at zero force.
\end{abstract}



\maketitle

\section{Introduction}
Single molecule force-measuring techniques have made possible the
controlled ma\-nip\-u\-la\-tion of individual molecules by applying
forces on the piconewton scale. Current force measuring devices
include the atomic force microscope, optical and magnetic
tweezers, or even microneedles and biological membranes used as force
probes. Single molecule manipulation has been used to investigate many
problems in molecular and cellular biophysics (see
\cite{Ritort06,Arias07} for recent reviews). To cite just a few
examples: the mechanical properties of biopolymers such as DNA
\cite{SmiCuiBus95,WanYinLanGelBlo97,DesMaiZhaPelBenCro02} have been
established for the first time; the folding/unfolding processes in
individual RNA or protein molecules \cite{RieGauOesFerGau97,Lip1,Fern}
have been investigated; the interactions between DNA and proteins
\cite{LegRobBouChaMar98}, but also between DNA and RNA, have been
studied at the molecular level \cite{YinWanSvoLanBloGel95}; the motion
of single molecular motors has also been followed in real
time\cite{FinSimSpu94,NojYasYosKin97,StrCroBen00,AbbGreShaLanBlo05,WenTin08}.

An important aspect of single molecule force experiments is the
possibility to monitor the time evolution of individual molecules by
recording the molecular extension. Measuring forces and extensions as
functions of time provides a lot of information about thermodynamics and
kinetics of individual molecules. A very useful technique to achieve this goal 
are optical tweezers, which are suited to accurately
measure forces in the range 0.1--100 pN. Using optical tweezers, it is
possible to derive the free energies of formation of biomolecules with
good accuracy. At the same time, optical tweezers allow us to
investigate questions related to the kinetics of folding, a challenging
problem in biophysics and statistical mechanics. A physical quantity
useful to characterize the behaviour of complex systems is the free
energy landscape. The free energy landscape describes the energetics of
the configurational space of a system. Introduced and applied in the
context of disordered and glassy systems, this concept finds a major
application in small systems, where thermal fluctuations entail large
conformational fluctations and the system can explore a large portion of
the configurational space. 

In this paper we investigate the thermodynamics and kinetics of force
induced folding/unfolding (hereafter referred as F/U) of short DNA
hairpins. Investigating force kinetics of DNA hairpins presents several
advantages \cite{WooAntBehLarHerBlo06,WooBehLarTraHer06} over 
other molecular constructs. In particular,
DNA sequences can be synthesized with relative ease and DNA degrades
comparatively less than other molecules (e.g. RNA) do.  We carry out
single DNA pulling experiments using optical tweezers to extract
information from thermodynamics and kinetics under the application of an external
force. The results are compared with theoretical predictions based on
the concept of the free energy landscape. The paper is divided as
follows. After a brief summary of the type of experiments in
 \sref{pulling}, we elaborate on the concept of free energy landscape
applied to nucleic acid hairpins (DNA or RNA) in
 \sref{free}.  \Sref{ruptureforce} explains how to extract
information about thermodynamics and kinetics from pulling data. An
analysis of the kinetic parameters extracted from our experiments is
presented in \sref{kineticpar}. \Sref{freeenergy} shows an
alternative method to derive free energy differences using fluctuation
relations. Finally, in \sref{dg0} we discuss how to extract the free
energy of formation of the hairpin at zero force, both from kinetics and
fluctuation relations. After the conclusive section, three appendices supplement
the results of this paper detailing some technical subjects.

\section{Mechanical unfolding of DNA hairpins}  
\label{pulling}

\begin{figure}
\begin{center}
\includegraphics[width=10cm,angle=90]{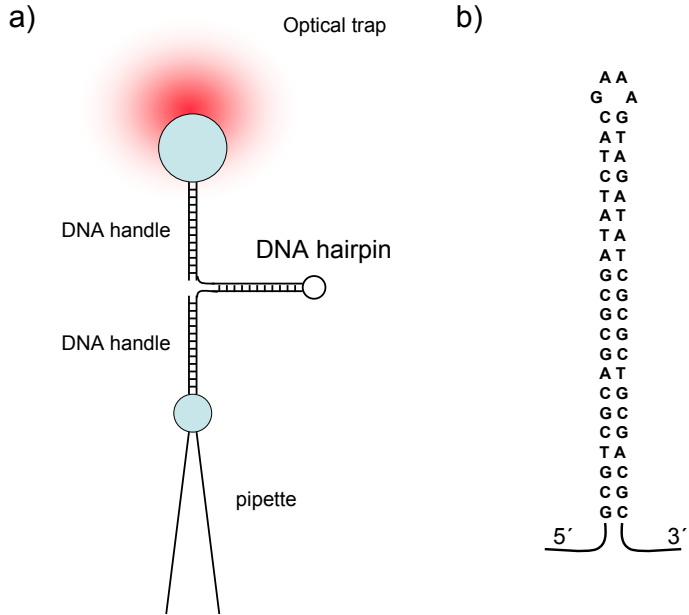}
\caption{(a) Experimental set up. (b) DNA hairpin sequence. The 5' and
  3' labels indicate the polarity of the phosphate chain of the hairpin.} 
\label{fig1}
\end{center}
\end{figure}

Our experimental setup is shown in \fref{fig1}(a). The DNA hairpin
is tethered between two beads by using double-stranded DNA (dsDNA) handles. Experiments are
carried out in a newly designed miniaturized dual-beam
laser optical tweezers apparatus \cite{Smith08}. One bead is immobilized in the tip of a
micropipette that is solidary with the fluidics chamber, the other bead
is captured in an optical trap generated by two counterpropagating laser
beams \cite{Smith1}. The force acting on the bead can be directly measured from the change 
in light momentum deflected by the bead. A steerable optical trap can be moved up and down along the vertical axis so to repeatedly unfold/refold the molecule. Every pulling cycle consists of a
stretching process (hereafter referred as S) and a releasing (hereafter
referred as R) process. In the stretching part of the
cycle the molecule is stretched from a minimum value of the
force ($f\sub{min}\sim 10$ pN), so small that the hairpin is always folded, up to
a maximum value of the force ($f\sub{max}\sim 20$ pN), so large that the hairpin
is always unfolded. During the releasing part of the cycle the force
is decreased from $f\sub{max}$ back to $f\sub{min}$. The force is
varied at the same {\it loading rate} (the rate at which the force is increased or decreased) in both stretching and releasing stages of the cycle\footnote{This process is performed at constant pulling speed $v$. Since the elasticity of DNA (dsDNA handles and unfolded DNA hairpin) is force-dependent, strictly speaking the loading/unloading rate $r$  is not constant throughout the pulling process.
Nevertheless, in the force range of our experiment the rigidity of the optical trap $k\sub{b}$ is much smaller than the stiffness of the handles and the unfolded DNA hairpin. Therefore, the effective rigidity \eref{keffect} is $k\sub{eff}\approx k\sub{b}$, and our system verifies $r=v k\sub{eff} \approx v k\sub{b}$.}, and recorded with an acquisition frequency of 1 kHz.  All experiments
were done at a temperature $23^{\circ}-24^{\circ}$ C in a 1M NaCl aqueous
buffer with neutral pH (7.5) stabilized by Tris HCl and 1M
EDTA.

The molecular construct is shown in \fref{fig1}(b) and consists of a
DNA hairpin of 21 base pairs (bps) ending with a tetraloop GAAA. The
hairpin is inserted between two identical short dsDNA handles of 29 bps
each. The sequence of this DNA hairpin
is canonical (i.e. all base pairs are complementary)
and has been specifically designed to produce a two-state folder (see below in \sref{free}).

Force-distance curves (FDCs) in our experiments represent the force acting on the molecule as a
function of the relative position of the trap along the force
axis. From the FDC it is possible to extract the molecular extension, so to
represent the force versus the molecular extension in what is known as a
force-extension curve (FEC). Many works in single molecule manipulation
often use such representation. However, in many aspects it is better to
use the trap position rather than the molecular extension to draw
pulling  curves. In fact, the former is the only parameter that is externally 
controlled (referred to as the \textit{control parameter} \cite{Ritort08}) whereas 
the latter is subject to fluctuations that  introduce additional (albeit small) corrections (e.g. in the
measure of the mechanical work exerted upon the molecule). Although either
description contain the same information, we will stick to the FDC picture
throughout this paper.

\Fref{fig5}(a) shows some typical FDCs for the sequence under study at
three different loading rates (slow and fast).  The FDC shows a linear
dependence of the force versus distance as a consequence of our choice
of short handles. In fact, because the handles are very rigid
(when compared to the rigidity of the trap) the effective rigidity of the
system made of bead and handles is mostly determined by the constant rigidity
of the Hookean (i.e. linear) optical trap.  Force rips are generated
each time the molecule folds/unfolds: after unfolding, a segment of 46
nucleotides of ssDNA is released so the force suddenly drops as the
trapped bead relaxes toward the center of the optical trap; whereas
after refolding the closure of the DNA hairpin pulls the bead away
from the trap and the force increases.

When pulling at slow loading rates the molecule shows low hysteresis
in the value of the unfolding/refolding force. Moreover, the hairpin
can execute several F/U transitions during the stretching and
releasing stages of the cycle. In contrast, when pulling at fast
loading rates, the molecule shows larger hysteresis in the value of
the unfolding/refolding force, and multiple F/U transitions are
unlikely. How many transitions are observed and how much irreversible
are the stretching and releasing processes depends on how large is the
pulling rate $r$ compared to the typical folding/unfolding rate
\cite{ManMosForHugRit08}.

\section{The free energy landscape}  
\label{free}
To follow the dynamics of the folding/unfolding process in configuration
space is a formidable task, due to the large number of interacting 
degrees of freedom. However, the collective behaviour displayed 
by the F/U transition suggests a simplified approach: one can judiciously 
choose one collective \textit{reaction coordinate} and project the 
many-dimensional energy landscape (where we represent the energy
of each microscopic configuration) onto a one-dimensional 
\textit{free energy} profile, where each point stands for an ensemble of
configurations. One minimum of the free energy profile represents
the energetically favored folded state, another one  the entropically 
favored unfolded state. In between, we find information about how much 
energy is needed (in the form either of heat or work) to explore intermediate
states of the system. In this section we will focus on the case where 
pressure and temperature are constant, the applied force is the control parameter,
while other variables such as the molecular extension or the position of 
the trap fluctuate. The effect of the experimental setup on the free energy landscape is
discussed in \ref{apa} and \ref{apc}.

As reaction coordinate, we choose the number of open base-pairs $n$.
If $N$ is the number of base-pairs in the stem of the hairpin, and $L$ 
is the number of bases in the loop (so that the total number of bases in the 
hairpin is $2N+L$), the configuration with $n=0$ is the folded state, while the 
configuration with $n=N$ is the completely unfolded state. In the absence of 
an externally applied force, we will use the symbol $G^0(n)$ to denote the free energy
to be delivered to the molecule (in the form of heat) in order to break the
first $n$ base pairs. $G^0(n)$ can be measured in bulk experiments,
e.g. by calorimetry or UV absorbance: by melting oligonucleotides of
different lengths we can enforce the desired number $n$ of dissociated base
pairs. The free energies of dissociation of the different nearest
neighbour base pairs \cite{SantaLucia1998} as well as other different
secondary structural elements (e.g. base-pair mismatches, loops, etc.) are
used by Mfold \cite{Zuker2003} to extract the free energy of formation of the DNA
molecule.  In general, $G^0(n)$ will be a monotonically increasing function
of $n$ whenever base pairs dissociate; however, it may decrease in
the presence of entropic structural elements such as loops \cite{SanHic04}.

\begin{figure}
\begin{center}
\includegraphics[width=14cm,angle=0]{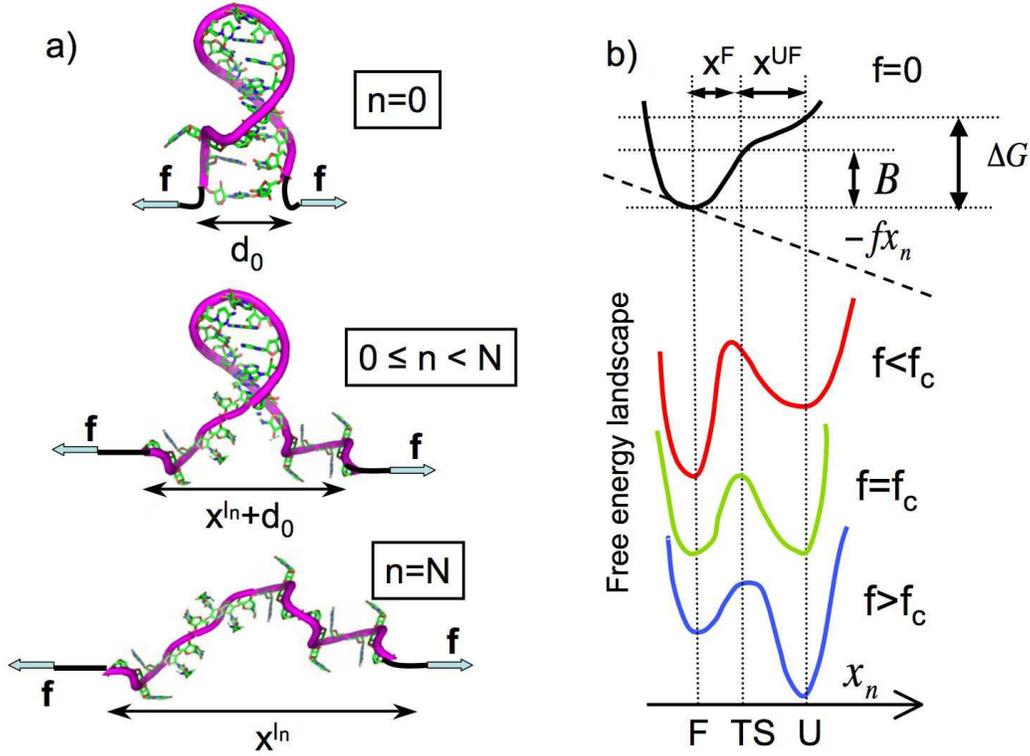}
\caption{(a) Schematic picture of the different configurations of a
  hairpin with $N$ base pairs at force $f$. For all configurations
  where the molecule is not completely unfolded, the molecular
  extension is equal to $x^{l_n}+d_0$ where $x^{l_n}$ is the extension
  at force $f$ of the released $2n$ bases of ssDNA. (b) Schematic
  picture of the free energy landscape at different forces $f$. The
  most important parameters are the distances from the transition
  state (TS) to the folded (F) and unfolded (U) states: $x^{\rm F}$
  and $x^{\rm UF}$, respectively; the free energy difference $\Delta
  G(f)$ and the barrier $B(f)$. As force increases the free energy
  landscape is tilted down favouring configurations of large extension
  $x_n$.}
\label{fig2}
\end{center}
\end{figure}

\subsection{The effect of force}
\label{free_ene}
What is the effect of an externally applied force $f$ on the free energy
landscape $G^0(n)$? To answer this question, we introduce the
force-dependent free energy landscape $G(x_n,f)$, where $x_n$ is the
distance between the 3' and the 5' extremities of the hairpin, measured along 
the direction of the applied force \cite{CocMonMar01,CocMonMar02}. 
We find it convenient to express the free energy in terms of  $x_n$ 
because this quantity, unlike $n$, is experimentally accessible.

In the folded state $n=0$, in the presence of an external force $f\neq 0$, the 
hairpin is always oriented along the force axis, therefore $x_0$ is equal to the 
diameter of the hairpin $d_0$ (typically about 2 nm). If $f = 0$, then the hairpin 
is generally not aligned, and all we can say is that $x_0 \leq d_0$. For any 
intermediate state $0<n<N$, $x_n$ is equal to $d_0$, plus the equilibrium
extension $x^{l_n}(f)$ of a ssDNA molecule with contour length $l_n=2nd$ (this relation 
can change in the presence of structural motifs such as base-pair mismatches) 
if $d$ is the inter-base distance. 
The length $x^{l_n}(f)$ is found by inverting the thermodynamic force-extension curve 
\be
	f=F_{\rm ssDNA}^{l_n}(x) \,.	\label{equaciof}
\ee
The explicit form of \eref{equaciof} depends on the particular model assumed (the most popular being the freely-jointed chain and the worm-like chain). 
In the totally unfolded state $n = N$, $x_N$ is the equilibrium extension $x^{l_N}(f)$
of a ssDNA molecule with contour length $l_N=(2N+L)d$. In synthesis, 
\be \label{xn(f)}
	x_n(f)=(1-\delta_{n,N})d_0+x^{l_n}(f)\,, \qquad l_n=2nd+\delta_{n,N}Ld \,.
\ee
Note that in \eref{equaciof} we assume thermodynamic equilibrium for the
ssDNA. It would be possible to include
elastic free energy fluctuations in the stretching part of the free
energy corresponding to the ssDNA to substitute the discrete $x_n$ with 
a continuous variable $x$, but this would not change much the 
predictions of the model.

 In the presence of an external force the free energy landscape $G(x_n,f)$ is tilted
along the reaction coordinate: 
\be
	G(x_n,f)=G^0(n)+G_{\rm ssDNA}(0\to x^{l_n};f)-fx_n \,,
	\label{gxf}
\ee
where we have introduced
\be
	G_{\rm ssDNA}(0\to x^{l_n};f)=\int_{0}^{x^{l_n}}F_{\rm ssDNA}^{l_n}(y)\rmd y \,,
	\label{work_stretch}
\ee
that is the reversible work needed to stretch a segment of ssDNA 
of contour length $l_n$ from an initial extension equal to
0 until a final extension $x^{l_n}(f)$ (see \fref{fig2}(a)), or, in other words, the area under the
force-extension curve \eref{equaciof} between 0 and $x^{l_n}$. 

Using \eref{work_stretch}, after a change of integration variable, we can rewrite
the free energy landscape \eref{gxf} as
\be
	G(x_n,f)=G^0(n)-\int_{0}^f x_n(f')\rmd f'+\delta_{n,N}fd_0
	\label{gxf2} \,.
\ee
From the free energy landscape \eref{gxf} we can define two important
force dependent parameters. The first is the free energy difference between the folded and the
unfolded states,
\be
	\fl \Delta G(f)=G(x_{N},f)-G(x_0,f)=\Delta G_0+G_{\rm ssDNA}(0\to x^{l_{N}};f)-(x^{l_{N}}-d_0)f\,,
	\label{dgfold}
\ee
where $\Delta G_0=\Delta G(f=0)=G^0(N)-G^0(0)$ denotes the free energy
of formation of the hairpin at zero force. In what follows we will simplify the notation
by defining
\be
	x\sub{m}(f)=x^{l_{N}}(f)-d_0\,.
	\label{notacioxm}
\ee
$x\sub{m}$ depends on the force $f$ and is equal to the released molecular
extension when the hairpin unzips completely. In this notation we have
\be
	\fl \Delta G(f)=G(x_{N},f)-G(x_0,f)=\Delta G_0+G_{\rm ssDNA}(0\to x^{l_{N}};f)-fx\sub{m}\,.
	\label{dgf}
\ee
The other important parameter is the free energy barrier, equal to the
free energy difference between the transition and the folded states,
\be
	B(f)=G(x^{\rm F},f)-G(x_0,f)=B_0+G_{\rm ssDNA}(0\to x^{l_{\rm F}};f)-fx^{\rm F}\,,
	\label{dbf}
\ee
where $x^{\rm F}$ is defined as the value of $x_n$ where $G(x_n,f)$ is
maximum, $B_0=G^0(n^{\rm F})-G^0(0)$ is the free energy barrier at
zero force, $n^{\rm F}$ is the number of base pairs released at the transition state at force $f$, and $l\sub{F}\equiv l_{n^{\rm F}}$. 

We stress that the parameters $\Delta G(f)$ and $B(f)$, as determined
from the shape of the free energy landscape $G(x_n,f)$, depend directly on
the applied force. Moreover, the relevant distances $x\sub{m}$ and $x^{\rm{F}}$ 
 are also expected to depend on the force due to the
dependence of the molecular extension of the ssDNA with force, as shown
in \eref{equaciof}. Yet, such dependence is expected to be
very small for forces in the vicinity of the F/U transition
(see below). In what follows, and in order to lighten the notation, we
will not indicate explicitly such force dependence for these
distances. For later use we define the following quantities:
\be
	\fl \Delta G_1(f) &= \Delta G_0+G_{\rm ssDNA}(0\to x^{l_{N}};f) &\Longrightarrow \Delta G(f)=\Delta G_1(f)-fx_m \label{g1a} \\ 
	 \fl B_1(f) &= B_0+G_{\rm ssDNA}(0\to x^{l\sub{F}};f) &\Longrightarrow B(f)=B_1(f)-fx^{\rm F} \,. \label{g2a}
\ee
As we will see below, $\Delta G_1$, $x\sub{m}$ and $x^{\rm F}$ can be directly measured in pulling experiments.

In the previous computation, we have neglected the work needed to orient the
hairpin along the force axis. By treating the
  elastic response of the folded hairpin as a polymer of contour length
  equal to the hairpin diameter $d_0$ and persistence length $P$, the
  work necessary to orient the hairpin along the force axis is inversely
  proportional to $P$. For a rigid object $P$ is large, so the free
  energy of orienting the hairpin is indeed expected to be negligible. Not so
  the effect of the hairpin diameter, as showed in \eref{gxf2}.

\subsection{Kramers--Bell kinetic rates}
\label{kramersbell}
According to the Kramers--Bell theory \cite{Bell78}, the kinetic rates of unfolding and
folding under tension in a two-state folder are given by
\numparts
\be
k_{\rightarrow}(f)&=k_0\exp{\left[-\left(\frac{G(x^{\rm F},f)}{k\sub{B}T}\right)\right]}\,,\label{rate_0a}\\
k_{\leftarrow}(f)&=k_0\exp{\left[-\left(\frac{-G(x_{N},f)+G(x_0,f)+G(x^{\rm F},f)}{k\sub{B}T}\right)\right]}\,.\label{rate_0b}
\ee
\endnumparts
The $k_{\rightarrow}(f)$ ($k_{\leftarrow}(f)$) describes the kinetic
rate to jump over the transition state from the folded (unfolded) state. $k_0$ stands for the
attempt frequency of the hairpin (which may get
contributions from the instrument and the whole molecular construct),
$k\sub{B}$ and $T$ being respectively the Boltzmann constant and the
temperature of the bath. The rates \eref{rate_0a} and \eref{rate_0b} satisfy detailed balance, 
\be
\fl \frac{k_{\rightarrow}(f)}{k_{\leftarrow}(f)}&=\exp{\left(-\frac{G(x_{N},f)-G(x_0,f)}{k\sub{B}T} \right)}\nonumber\\
\fl &=\exp{\left(-\frac{\Delta
G_0+G_{\rm ssDNA}(0\to x^{l_{N}};f)-fx\sub{m}}{k\sub{B}T} \right)}=\exp{\left(-\frac{\Delta G(f)}{k\sub{B}T} \right)} \,,
\label{db}
\ee
where $\Delta G(f)$ has been defined in \eref{dgf}. The coexistence force $f\sub{c}$ is defined
as the value of the force at which the equilibrium constant is equal to one,
\be
k\sub{c}=k_{\rightarrow}(f\sub{c})=k_{\leftarrow}(f\sub{c}) \,,
\label{deffc}
\ee
$k\sub{c}$ being the coexistence rate. According to \eref{db}, this corresponds to $\Delta G(f\sub{c})=0$, or 
\be
 \Delta G_1(f\sub{c})=\Delta G_0+G_{{\rm ssDNA}}(0\to x^{l_{N}};f\sub{c})=f\sub{c}x\sub{m} \,.
\label{critf}
\ee
Summarizing, the kinetic rates \eref{rate_0a} and \eref{rate_0b} can be rewritten
in a more compact form,
\numparts
\be
k_{\rightarrow}(f)&=k_{0}\exp{\left[-\left(\frac{B_1(f)-fx^{\rm{F}}}{k\sub{B}T}\right)\right]}\,,\label{rate_2sa}\\
k_{\leftarrow}(f)&=k_{0}\exp{\left[-\left(\frac{B_1(f)-\Delta
G_1(f)+fx^{\rm{UF}}}{k\sub{B}T}\right)\right]}\,,
\label{rate_2sb}
\ee
\endnumparts
where 
\be
x\sub{m}=x^{\rm F}+x^{\rm UF}
\label{xm}
\ee
is equal to the change in molecular
extension across the F/U transition. The terms $\Delta G_1(f),B_1(f)$
are given in \eref{g1a} and \eref{g2a}, while each stretching contribution (contained
in those terms) of the type $G_{{\rm ssDNA}}(0\to x^l(f))$ has been defined
in \eref{work_stretch} and  \eref{equaciof}. 

\subsection{Simplified version of the rates}
\label{simplified}

\begin{figure}
\vspace{0.9cm}
\begin{center}\includegraphics[scale=0.4,angle=0]{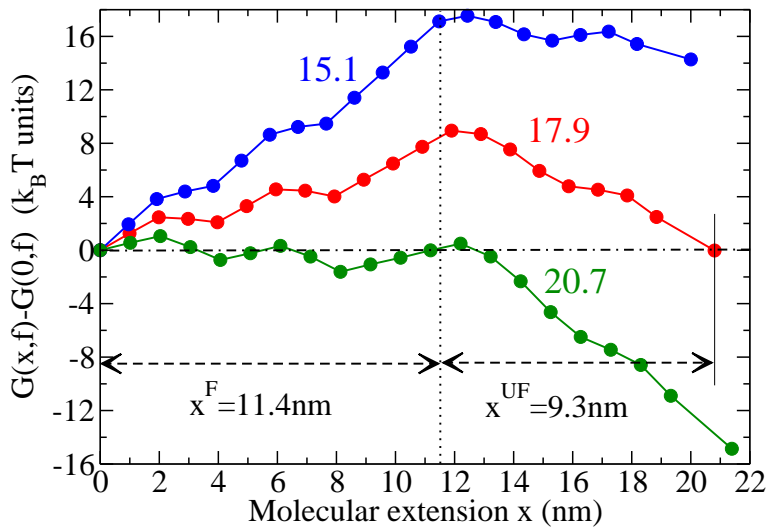}
\caption{ Free energy landscape at various forces for the hairpin
  sequence shown in \fref{fig1}(b). The stretching contribution has been
  calculated from Mfold using \eref{dg01} and \eref{dg02} (see below in
  \sref{dg0}). Moreover, the effect of the diameter $d_0$ has been included in
  the calculation. The values shown for the distances $x^{\rm
    F},x^{\rm UF}$ are approximately 1 nm larger than the values found
in our experiments. However, the value for the fragility $\mu$ \eref{mu} inferred
from these distances (0.1) is compatible with that  found in the experiments
(0.0975, see \tref{table1}).}
\label{fig3} 
\end{center} 
\end{figure} 

In a further simplified description \cite{Tinoco03}, it is common to neglect any force
dependence for $\Delta G_1(f)$ and $B_1(f)$ in
\eref{rate_2sa} and \eref{rate_2sb}, so to incorporate these quantitites into 
effective values for the barrier $B_1$ and the free energy difference
$\Delta G_1$ measured at the transition force $f\sub{c}$. This means that $\Delta
G_1(f)\simeq \Delta G_1(f\sub{c})$ and $B_1(f)\simeq B_1(f\sub{c})$ if $f$ is not
too far from $f\sub{c}$. This approximation is justified because the elastic
contributions contained in those terms vary much less with force than the products $fx^{\rm{F}},fx^{\rm{UF}}$
do. A mathematical proof of this result is shown in
\ref{apb}. In what follows we will drop any force dependence in the effective
parameters $B_1$ and
$\Delta G_1$ and write the kinetic rates as follows:
\be
k_{\rightarrow}(f)&=k\sub{m}\exp{\left(\frac{f x^{\rm F}}{k\sub{B}T}\right)}\,,
\qquad k_{\leftarrow}(f)=k\sub{m}\exp{\left(\frac{\Delta G_1-fx^{\rm UF}}{k\sub{B}T}\right)}\,,
\label{e7}
\ee
where $k\sub{m}$ corresponds to the
unfolding rate at zero force and is given by 
\be
k\sub{m}=k_0\exp{\left(-\frac{B_1}{k\sub{B}T}\right)}\,. 
\label{km}
\ee
As we will see in \sref{ruptureforce}, the simplified rates \eref{e7} reproduce reasonably
well the experimental results.

The success of the two-states model in reproducing the kinetics of the
force induced F/U reactions depends on the hairpin
sequence. In the absence of structural motifs that may induce
alternative F/U pathways or intermediate states (see for example \cite{ManJunRit09}), a short
hairpin (a few tens of base pairs) will display
cooperative two-states behaviour. In addition, if the sequence is
designed in such a way that the free energy landscape along the
reaction coordinate has a single barrier, then such molecule is
expected to behave as a two-state folder displaying simple Arrhenius
kinetics. In this work we have designed a DNA sequence (see
\fref{fig1}(b)) that has such properties. The free energy landscape as function of the
molecular extension has been calculated at various forces using the
free energy values from Mfold and the elastic properties of ssDNA. In \fref{fig3} we
show the calculated free energy landscape at various forces around
the coexistence force $f\sub{c}\simeq 17.9$ pN, where the free
energies of the folded and unfolded states are equal.

\subsection{The fragility}

An important aspect of the Kramers--Bell rates \eref{e7} is their strong
dependence on force which is determined by the values of $x^{\rm F}$ and $x^{\rm
UF}$. If $x^{\rm F}\gg x^{\rm UF}$, then the transition state is located far
away from the folded state and the molecule deforms
considerably before it unfolds. In the other case, when $x^{\rm UF}\gg
x^{\rm F}$, the transition state is located close to the folded state
and the molecule unfolds without deforming much. 
A quantitative measure of how much the native structure deforms before unfolding
occurs is given by the fragility parameter \cite{Lef53,HyeThir05,HyeThir06,man1}
\be
\mu=\frac{x^{\rm F}-x^{\rm UF}}{x^{\rm F}+x^{\rm UF}}=\frac{x^{\rm F}-x^{\rm UF}}{x\sub{m}}\,.
\label{mu}
\ee
$\mu$ lies in the range $[-1:1]$ and defines the degree of compliance of
the molecule under the effect of tension. Fragile or compliant molecules are those in which
$x^{\rm F}$ is larger than $x^{\rm UF}$ and $\mu$ is positive. In
contrast, when $x^{\rm UF}$ is larger than $x^{\rm F}$ and $\mu$ is
negative, we talk about brittle structures.  The fragility has been proved 
to be a useful parameter to describe the mechanical unfolding of RNA
hairpins with more than one transition state \cite{man1}. 

\section{Breakage force kinetics}
\label{ruptureforce}

In order to manipulate a DNA hairpin using optical tweezers, the free
ends of the molecule are attached to micron-sized beads by using dsDNA
handles. In this experimental configuration (see
\fref{fig1}(a)), the force fluctuates and the control parameter is the
extension between the center of the trap and the tip of the
micropipette. This experimental setup corresponds to the so called
mixed ensemble.  It is then possible to describe the F/U kinetics of
the DNA hairpin using the two-state model depicted in \fref{fig2},
taking into account that the variable that controls the shape of the
free energy landscape is the trap--pipette distance rather than the
force. The mixed ensemble gets contributions from the different
elements of the experimental setup. In their simplified form
the kinetic rates in the mixed ensemble can be shown to obey \eref{e7}
with identical force dependent terms in the exponent (i.e. equal
values for $\Delta G_1,x^{\rm F},x^{\rm UF}$) but different prefactors
(see Appendix C in \cite{ManRit05}).

\begin{figure}
\begin{center}
\vspace{0.9cm}
\includegraphics[scale=.4,angle=0]{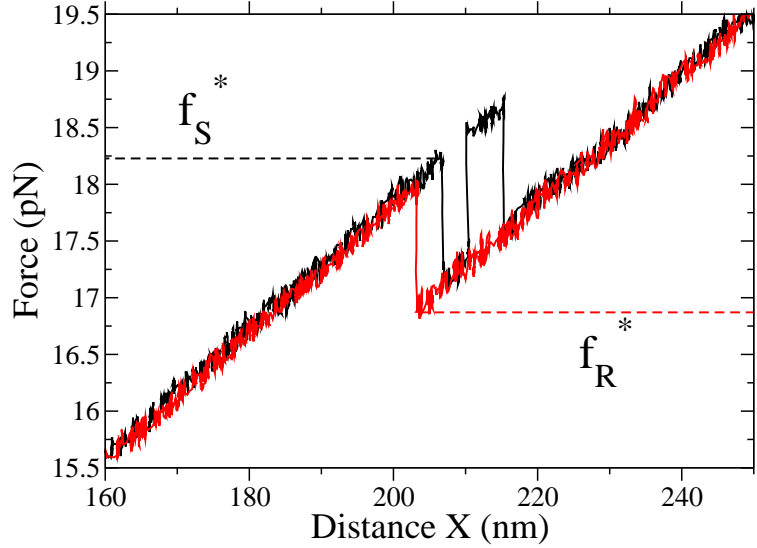}\
\caption{ A representative FDC showing the stretching (black) and releasing (red)
parts of a pulling cycle. It shows the first rupture force during stretching ($f^*\sub{S}$) and releasing
($f^*\sub{R}$) processes.} 
\label{fig4}
\end{center}
\end{figure}

 The simplest way to extract the values of $\Delta G_1,x^{\rm{F}},x^{\rm{UF}}$
from the pulling data is to analyze the distribution of first rupture
forces along stretching and releasing parts of the cycle, $f^*_{\rm{S}}$
and $f^*_{\rm{R}}$. The first rupture force $f^*_{\rm{S(R)}}$ along the
stretching (releasing) part of the cycle is the value of the force at
which the first force rip is observed at the time where the first jump
occurs. An illustration is shown in \fref{fig4}. Useful quantities that can be measured
in pulling experiments are: the survival probability $P_{\rm S(R)}(f)$ that
the molecule remains in the folded (unfolded) state along the stretching (releasing)
process until reaching the force $f$, the mean value and the variance of
the first rupture forces $f^*_{\rm{S(R)}}$.

\begin{figure}
\begin{center}
\includegraphics[scale=.29,angle=0]{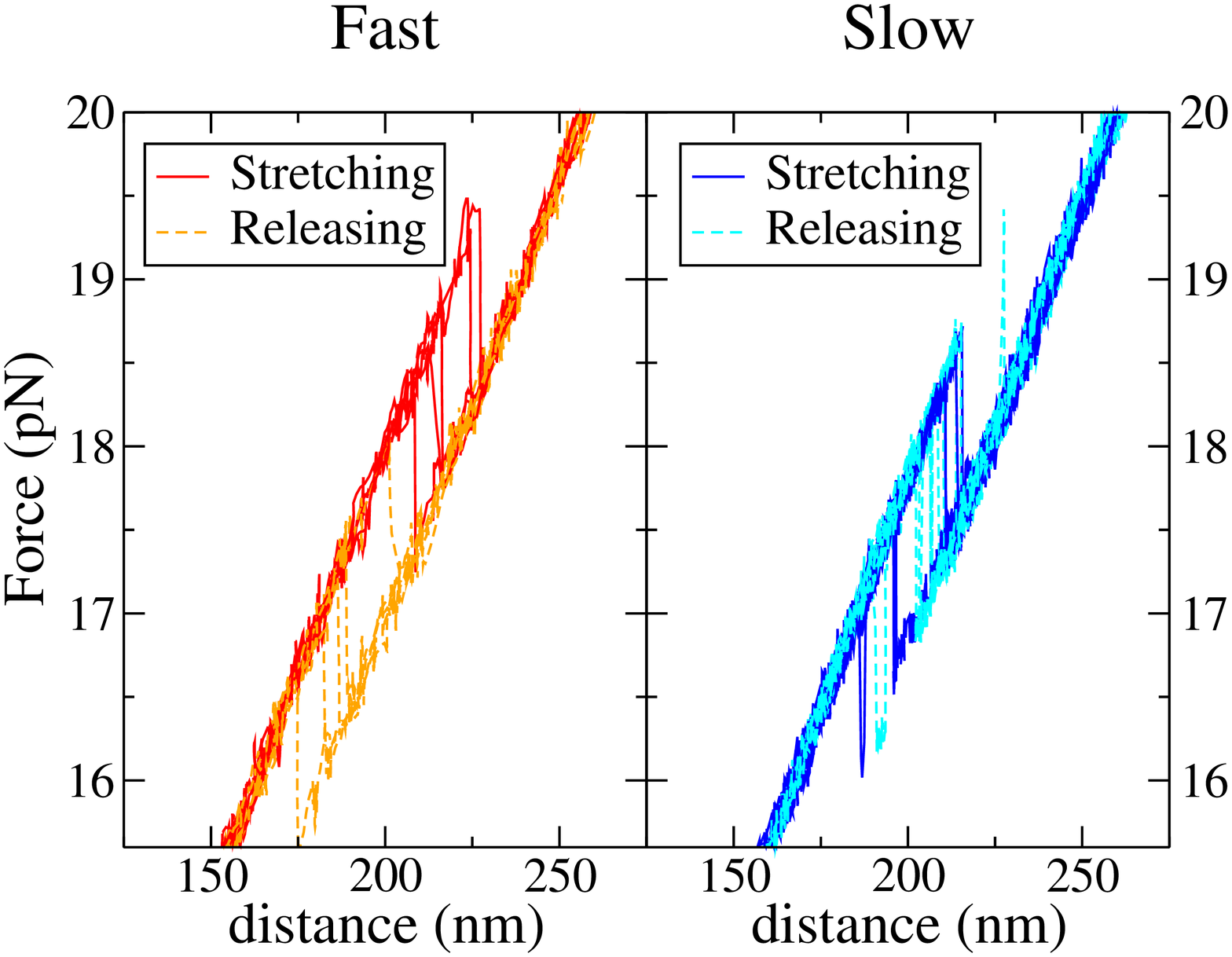}\hspace{0cm}\includegraphics[scale=.29,angle=0]{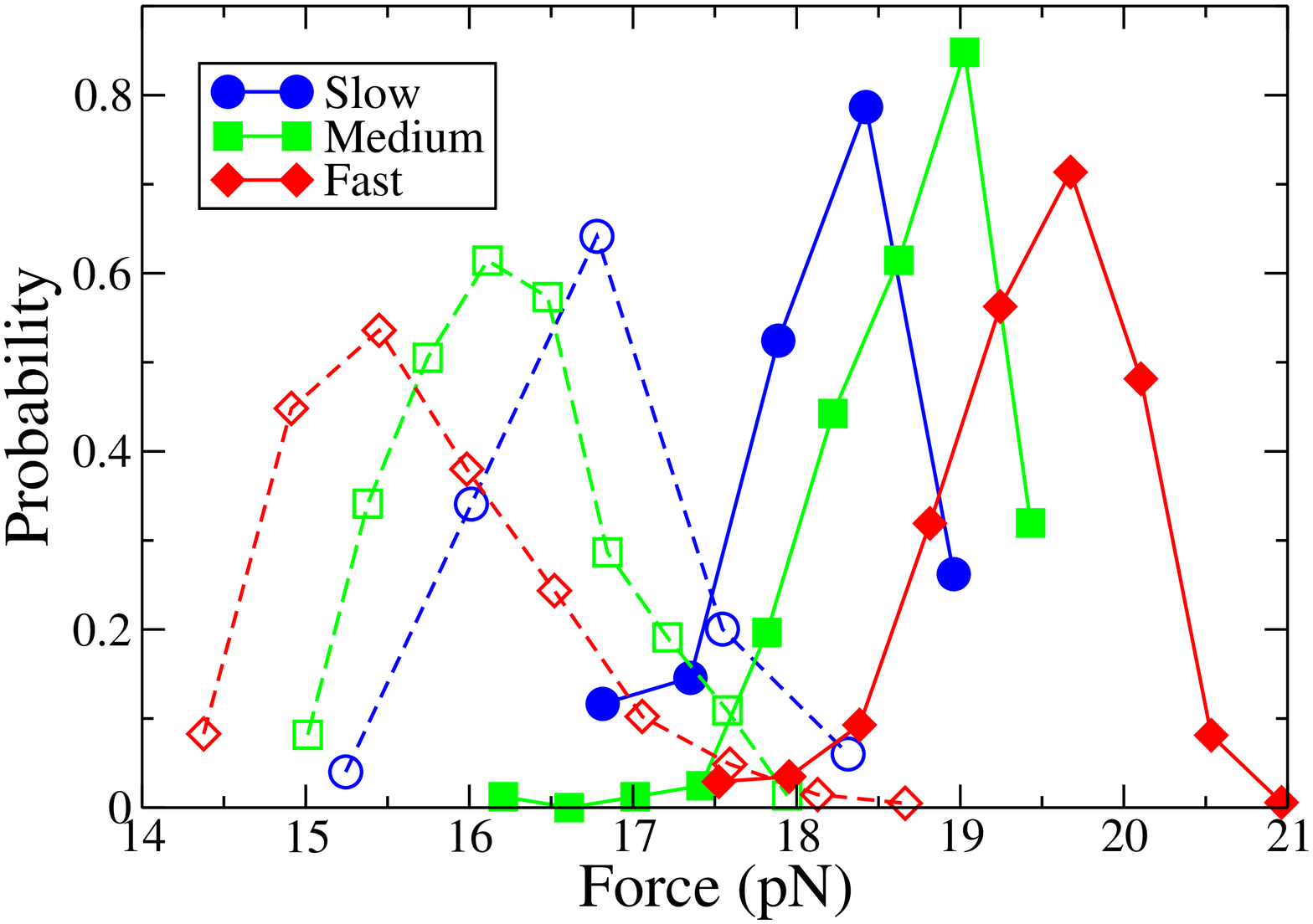}
\caption{ (a) Force-distance curves (FDCs) corresponding to 5 cycles
  at fast and slow pulling speeds (200 nm/s and 25 nm/s respectively,
  corresponding to loading rates of 8.1 and 1.0
  pN/s). Continuous red (dashed orange) lines correspond to fast
  stretching (releasing) parts of the cycle. Continuous blue (dashed
  cyan) lines correspond to slow stretching (releasing) parts of the
  cycle.  Note that hysteresis increases with the pulling
  speed. Moreover, FDCs at slow speeds show several transitions
  between the folded and the unfolded states. (b) First
  rupture force distributions for the stretching (continuous lines) and
releasing process (dashed lines) for three molecules
  at three different loading rates $r=1.76$ pN/s (blue), $r=4.8$ pN/s
  (green), $r=14.5$ pN/s (red).}
\label{fig5}
\end{center}
\end{figure}

\subsubsection*{Survival probability.} 
\begin{figure}
\begin{center}
\vspace{0.9cm}
\includegraphics[width=10cm,angle=0]{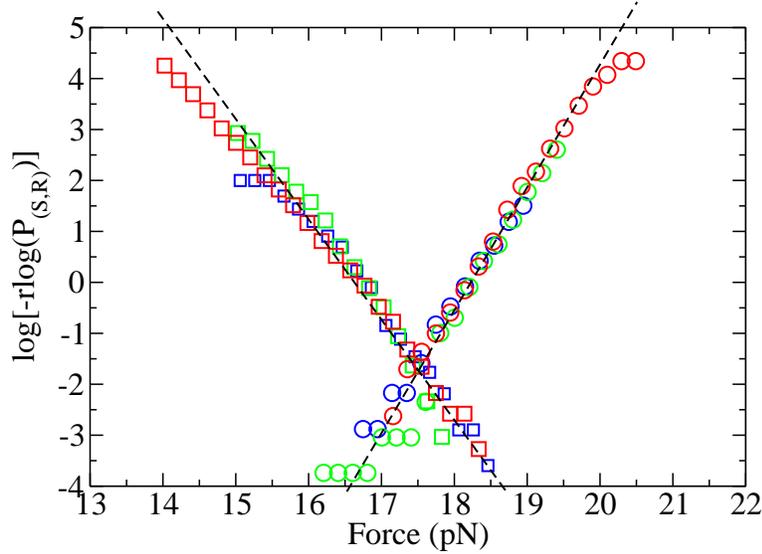}
\caption{ Experimental data for $\log(-r\log(P_{\rm (R,S)}))$ plotted
  as function of force can be fit to straight lines according to \eref{logpu}, \eref{logpr}. The slopes
  of the lines give the distances $x^{\rm F}=9.8\ \mathrm{nm},x^{\rm
    UF}=8.1\ \mathrm{nm}$ (see \tref{table1}). Data correspond to three
  different molecules pulled at different loading rates $r=1.76$ pN/s 
  (blue), $r=4.8$ pN/s (green), $r=14.5$ pN/s (red). Circles correspond to
  the stretching process. Squares correspond to the releasing
  process. Note that the value of the force at which folding and
  unfolding lines cross each other ($\simeq 17.5$ pN) lies pretty close
  to the average coexistence force, 17.75 pN (see \tref{table1}).}
\label{fig6}
\end{center}
\end{figure}

The distribution $P_{\rm S(R)}(f)$ satisfies the following master equation:
\be
\frac{\rmd P_{\rm S(R)}(f(t))}{\rmd t}=-k_{{\rightarrow(\leftarrow)}}(f(t))P_{\rm S(R)}(f(t))\,.
\label{rur}
\ee  
$P_{\rm S(R)}(f)$ is related to the experimentally measured 
distribution of first rupture forces, $\rho_{\rm S(R)}(f)$, by
\be
P_{\rm S}(f)=1-\int_{f_{\rm min}}^{f}\rho_{\rm S}(f')\rmd f'\,,\qquad P_{\rm
  R}(f)=1-\int_{f}^{f_{\rm max}}\rho_{\rm R}(f')\rmd f' \,,
\label{int_surv}
\ee
 where $f_{\rm min}$ ($f_{\rm max}$) is the initial (final) force along the
 stretching-releasing cycles. For a protocol at a constant
 loading/unloading rate  $r$, $P_{\rm S(R)}(f)$ can be exactly computed:
\numparts
\be 
P_{\rm S}(f)&=
\exp{\left[-\left(\frac{k\sub{B}T}{rx^{\rm{F}}}\left(k_{\rightarrow}(f)-k_{\rightarrow}(f_{\rm min})\right)\right)\right]} \,,  \\
P_{\rm R}(f)&=
\exp{\left[-\left(\frac{k\sub{B}T}{rx^{\rm{UF}}}\left(k_{\leftarrow}(f)-k_{\leftarrow}(f_{\rm max})\right)\right)\right]} \,,
\label{rur2}
\ee
\endnumparts
where we used the unfolding (folding) rates given in \eref{e7}
\cite{Ev}. In the limit $f_{\rm min}\ll f\sub{c}$ (S process) and $f_{\rm max}\ll f\sub{c}$ (R
process), the function $\log [-r(\log(P_{\rm{S(R)}}(f))]$ is given by
\numparts
\be
\log [-r(\log(P_{\rm S}(f))]&=\log\left(\frac{k\sub{B}Tk\sub{m}}{x^{\rm F}}\right)+\left(\frac{x^{\rm
F}}{k\sub{B}T}\right)f \,, \label{logpu}\\
\log [-r(\log(P_{\rm R}(f))]&=\log\left(\frac{k\sub{B}Tk\sub{m}}{x^{\rm UF}}\right)+\frac{\Delta
G_1}{k\sub{B}T}-\left(\frac{x^{\rm UF}}{k\sub{B}T}\right)f \,. \label{logpr}
\ee
\endnumparts
These results show that  the function $\log [-r(\log(P_{\rm{S(R)}}(f))]$,
plotted as a function of the applied
force $f$, is a straight line with a slope inversely
proportional to the position of the kinetic barrier $x^{\rm{F}}$
($x^{\rm{UF}}$) and intercepts ($a_{\rm
S},a_{\rm R})$  related to the rate $k\sub{m}$ and the free energy difference $\Delta G_1$
\cite{Ev}. To extract the value of $\Delta G_1$, we do linear fits to
\eref{logpu} and  \eref{logpr}, then use the independent coefficients $a_{\rm
S},a_{\rm R}$ to obtain
\be
\frac{\Delta G_1}{k\sub{B}T}=a_{\rm R}-a_{\rm S}+\log\left(\frac{x^{\rm{UF}}}{x^{\rm{F}}}\right)\,.
\label{deltag1}
\ee
In order to extract $B_1$ we can use the relation \eref{km},
$B_1=-k\sub{B}T\log(k\sub{m}/k_0)$ where $k\sub{m}$ can be extracted from the value of
$a_{\rm S}$ but $k_0$ is unknown. In order to determine $B_1$ one needs
to make further assumptions. For example, it is possible to use effective
one-dimensional Kramers models applied to molecular free energy
landscapes \cite{man1,SchRie06,ShiSei96} to infer the value of $B_1$. \Fref{fig6} shows an experimental test of \eref{logpu}, \eref{logpr} using the results shown in \fref{fig5}(b).

\subsubsection*{Mean and variance of first rupture forces.} 
\begin{figure}
\begin{center}
\vspace{0.9cm}
\includegraphics[width=10cm,angle=0]{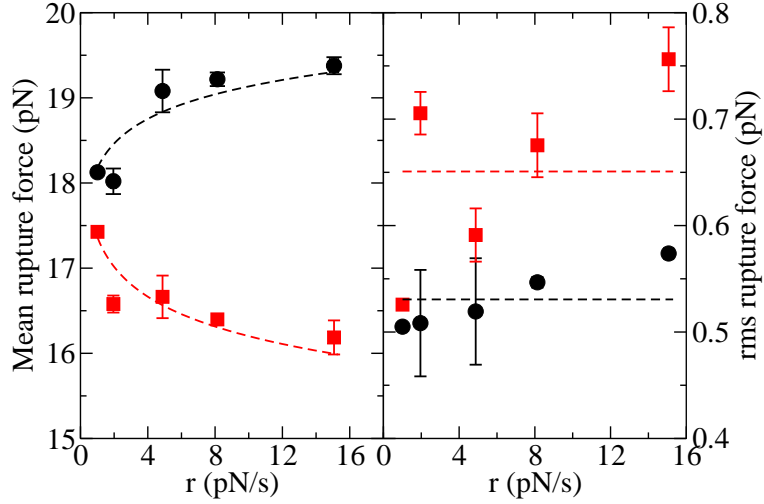}
\caption{ Average and rms deviation of first rupture forces as a function of
  the loading rate. Data have been taken by averaging over different
  molecules at different loading rates (see \tref{table1}). The average
  rupture force (left panel) has been fit to \eref{fa} fixing the values
  $x^{\rm F}=9.8\ \mathrm{nm},x^{\rm UF}=8.1\ \mathrm{nm}$ (\tref{table1}). We get $C=43.64,
  C'=34.35$ for the best fit. The rms of the rupture force (right
  panel) is constant with the loading rate \eref{fb} but cannot be fit to \eref{fb} with the values of $x^{\rm
    F},x^{\rm UF}$ obtained from the survival probability analysis. Instead we get $x^{\rm F}=7.74\ \mathrm{nm},x^{\rm
    UF}=6.34\ \mathrm{nm}$ (again compatible with $\mu=0.1$), which however are
  30\% smaller than the expected values.}
\label{fig7}
\end{center}
\end{figure}

The dependence on the rate $r$ 
of the mean value and the standard deviations of $f^*_{\rm{S}}$ and 
$f^*_{\rm{R}}$  can be computed in the relevant experimental regime, that is when
\be
a\equiv\frac{k_{{\rightarrow(\leftarrow)}}(f\sub{c})k\sub{B}T}{rx^{\rm{F(UF)}}}\ll 1\,,
\ee
($f\sub{c}$ being the coexistence force) \cite{Hum1}. The result is
\be
 \langle f^*_{\rm{S}}\rangle=\frac{k\sub{B}T}{x^{\rm{F}}}[C+\log(r)]+\Or(a)\,,\qquad
&\langle \sigma_{f^*_{\rm S}}\rangle =\frac{k\sub{B}T}{x^{\rm{F}}}+\Or(a)\,, \label{fa}\\
\langle f^*_{\rm{R}}\rangle=\frac{k\sub{B}T}{x^{\rm{UF}}}[C'-\log(r)]+\Or(a)\,, \qquad 
&\langle\sigma_{f^*_{\rm R}}\rangle=\frac{k\sub{B}T}{x^{\rm{UF}}}+\Or(a)\,, \label{fb}
\ee
where $\langle\sigma_{f^*_{\rm S}}\rangle,\langle\sigma_{f^*_{\rm
    R}}\rangle$ denote the rms deviation of the first rupture forces. The
constants $C$ and $C'$ depend on the characteristics of the molecule and
on the initial force values along the stretching and releasing processes
respectively. \Eref{logpu}, \eref{logpr} and \eref{fa}, \eref{fb} provide a way to extract the
relevant parameters that characterize the free energy landscape from the experimental
data. A test of the validity of \eref{fa}, \eref{fb} is shown in \fref{fig7}.

\section{Kinetic parameters for the hairpin}
\label{kineticpar}

In this section we use the already cited experimental data to extract
the parameters that characterize the free energy landscape and the
kinetics of the hairpin. A summary of the results obtained by analyzing
data for 11 molecules pulled at different speeds is shown in \tref{table1}. 
We have extracted the different values of the parameters for each
molecule to find the mean and the standard deviation. The values $x^{\rm F},x^{\rm UF}$ have been extracted from the linear fits \eref{logpu}, \eref{logpr}, whereas $x\sub{m}$ and $\mu$ are given by
\eref{xm},\eref{mu}. For each molecule the value of $\Delta G_1$ has been obtained from \eref{deltag1}
whereas the coexistence force $f\sub{c}$ and the coexistence rate $k\sub{c}$ are
extracted from \eref{critf}, \eref{deffc} and \eref{e7}:
\be
f\sub{c}=\frac{\Delta G_1}{x\sub{m}} \,, \qquad k\sub{c}=k\sub{m}\exp\left(\frac{f\sub{c}x^{\rm F}}{k\sub{B}T}\right) \,.
\label{fckc}
\ee
To complement such estimates we also show another estimate ($\overline{f\sub{c}}$) for the
average coexistence force which corresponds to the average first
rupture forces along the stretching and release parts of the cycle:
\be
\overline{f\sub{c}}=\frac{1}{2}(\overline{f^*_{\rm S}+f^*_{\rm R}})\,.
\label{avfc}
\ee
In addition, we also verify that the product of the coexistence force
$\overline{f\sub{c}}$ \eref{avfc} times the extension $x\sub{m}$ averaged over all molecules
is compatible with the average value of $\Delta G_1$. Finally, we also
show three more quantitites: 1) the average force jump across the
transition, $\Delta f$; 2) the average slope of the FDC corresponding to
the combined stiffness $k_{\rm eff}$ of bead and handles \cite{GerBunHwa01,GerBunHwa03},
\be
\frac{1}{k_{\rm eff}}=\frac{1}{k\sub{b}}+\frac{1}{k\sub{h}} \,;
\label{keffect}
\ee
and  3) the average retraction of the combined extension of the bead and
handles ($x\sub{b}+x\sub{h}$) induced by the force
change, $\Delta x=\Delta f/k_{\rm eff}$.
\Table{Mean and standard deviations of different parameters obtained
from the kinetic experiments. Averages are taken over 11 different
molecules that have been pulled at different
pulling speeds: 18.5 nm/s or 1 pN/s (1 molecule, 193 cycles), 36.5 nm/s or 1.95 pN/s (2
molecules,160 cycles), 86.2 nm/s or 4.88 pN/s (3 molecules, 570 cycles), 156 nm/s or 8.1 pN/s (2
molecules, 725 cycles), 274.3 nm/s or 14.9 pN/s (3 molecules, 1580
cycles).\label{table1}}
\br
$x^{\rm F}$ (nm)&$x^{\rm UF}$ (nm)&$x\sub{m}$ (nm)&$\mu$&$f\sub{c}$ (pN)&$\overline{f\sub{c}}$ (pN)\\
\mr
9.87(36)&8.13(33)&18.06(52)&0.097(23)&17.91(09)&17.75(08) \\
\br
$\Delta G_1$ ($k\sub{B}T$)&$\overline{f\sub{c}}\cdot x\sub{m}$ ($k\sub{B}T$)&$k\sub{c}$ (Hz)&$\Delta f$ (pN)&$k_{\rm eff}$ (Hz)&$\Delta x$ (nm)\\
\mr
78.7(2.4)&78.0(2.3) &0.58(08)&1.230(47)&0.0544(10)& 22.6(1.2)\\
\br
\endTable

From \tref{table1} it emerges a remarkable fact: the expected
molecular extension $x\sub{m}$ at $f_c=17.91pN$ is 3 nm smaller than the change in extension of
combined system formed by bead and handles, which is around 21.30 nm if we
subtract to the total extension of the ssDNA (around 23.30 nm, see below
in \tref{table2}) the diameter of the hairpin $d_0=2$ nm. Is this
expected? What is the real average change in molecular extension of the
hairpin across the F/U transition? The net change in molecular
extension across the transition can be estimated from the average
retraction experienced by bead and handles, $\Delta x=\Delta f/k_{\rm eff}$. 
  
  Assuming that handles have a large but finite stiffness (around
500 pN/$\mu$m), part of the retraction ($\simeq 10\%$ of the total
released molecular extension) might be accounted for by the retraction
of the handles. This makes $\Delta x\simeq 22.6$ (\tref{table1})
an upper bound to the released molecular extension by the ssDNA. 

\begin{figure}
\begin{center}
\vspace{0.9cm}
\includegraphics[width=10cm,angle=0]{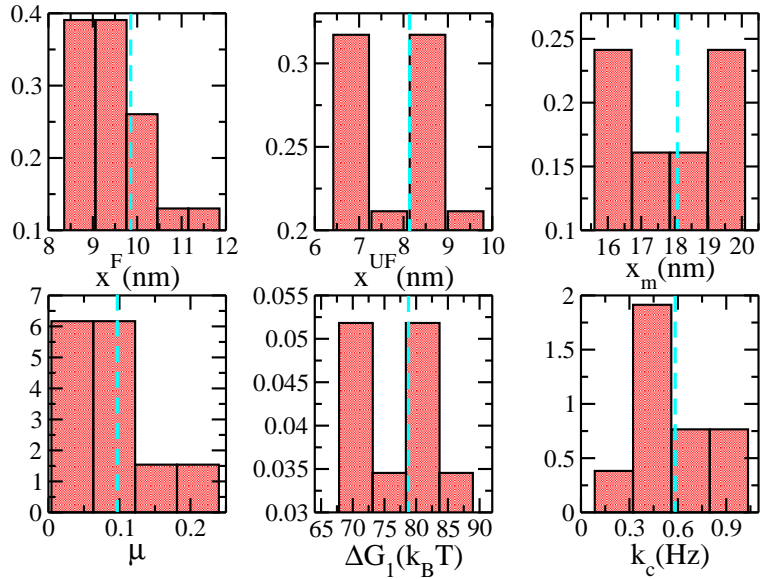}
\caption{  Histograms of some of the parameters reported in
  \tref{table1}. Statistics is collected over 11 molecules pulled
  at different spedds (see caption of \tref{table1} for more details). The
  vertical dashed lines (cyan color) show the mean of the distribution
  as given in \tref{table1}.}
\label{fig8}
\end{center}
\end{figure}

An important feature of our experiments
is the variability observed in the values measured for different
molecules. In \fref{fig8} we show histograms of values obtained for
a given set of parameters. Although distances $x^{\rm F},x^{\rm UF},x\sub{m}$
and the free energy value, $\Delta G_1$, typically show a 15$\%$ variation around the average value, other
quantitites like $\mu$ or $k\sub{c}$ show a more large variability.

Is it possible to infer the value $\Delta G_0$ from the
reported value for $\Delta G_1$? The inclusion of the stretching
contributions in order to infer the value of $\Delta G_0$ is discussed below
in \sref{dg0}.

\section{Free energy recovery}
\label{freeenergy}

Alternative methods to extract the free energy difference $\Delta G_1$
are provided by fluctuation relations. Fluctuation relations are
symmetry identities that relate the probability of a system to absorb or
release a given amount of energy to the environment along irreversible
processes. In our pulling experiments, single molecules are in a 
transient nonequilibrium state, as revealed by the systematic hysteresis observed
between the stretch and release processes at forces around the coexistence region. When the trap is moved fast
enough, then the free energy landscape is modified too quick and the
molecule cannot populate the different states (folded and unfolded)
according to the Boltzmann--Gibbs weight. Under equilibrium
conditions, \eref{db} predicts
\be
\frac{p_{\rm UF}(f)}{p_{\rm F}(f)}=\frac{k_{\rightarrow}(f)}{k_{\leftarrow}(f)}=\exp{\left(-\frac{\Delta G(f)}{k\sub{B}T} \right)} \,.
\label{db2}
\ee
Hysteresis effects indicate that measured populations of folded and
unfolded molecules $p_{\rm UF}(f),p_{\rm F}(f)$, averaged over many
pulling cycles, will not satisfy \eref{db2}. This precludes the validity of
thermodynamic
equilibrium in our pulling experiments. Note that mechanical equilibrium is probably satisfied
at the experimentally accessible pulling speeds, as revealed by the fact
that all other relaxational timescales in the system (beads, handles and
ssDNA) are much shorter than the F/U timescale \cite{WenManLiSmiBusRitTin07,ManWenLiSmiBusTinRit07,HyeMorThi08}.

\subsection{The Crooks fluctuation relation}
\label{crooks}
Let $W_{\rm S(R)}$ denote the mechanical work exerted on the molecule by
moving the optical trap. This is given by 
\be
W\sub{S(R)}=\int_{X_{\rm min}}^{X_{\rm max}}\rmd X F\sub{S(R)}(X) \,,
\label{intro1}
\ee
where the subindex S (R) refer to the stretching (releasing) stage
of the cycle.  In what follows, we will take $W\sub{S}$ and $W\sub{R}$ as
positive and negative quantities respectively, although, strictly speaking and according to
\eref{intro1}, both have positive signs. In fact, during the stretching (releasing) parts of the cycle the optical trap
delivers (extracts) mechanical work to (from) the molecule. During the
stretching part of the cycle the work is positive ($\rmd X>0$) whereas
during the releasing part of the cycle it is negative ($\rmd X<0$). By
repeatedly pulling the molecule many times we can measure the stretching
and releasing work
distributions,
\be
P\sub{S}(W)=\langle \delta(W-W_{\rm S})\rangle\,,\qquad P\sub{R}(W)=\langle \delta(W-W_{\rm R})\rangle\,.
\label{workprob}
\ee
where $\langle ..\rangle$ stands for the average over trajectories. For
an infinite number of pulls, the Crooks fluctuation relation
\cite{Crooks99} establishes that the probability distributions
\eref{workprob} satisfy the following identity:
\be
\frac{P\sub{S}(W)}{P\sub{R}(-W)}=\exp{\left( \frac{W-\Delta G_{X_{\rm min}}^{X_{\rm max}}}{k\sub{B}T}\right)}\,.
\label{crooksfr}
\ee
This relation provides an experimental way to extract the value of
$\Delta G_{X_{\rm min}}^{X_{\rm max}}$ from measurements of the
irreversible work. In particular, the two distributions $P\sub{S}(W)$ and
$P\sub{R}(-W)$ cross each other at the reversible work value, $W=\Delta
G_{X_{\rm min}}^{X_{\rm max}}$, thus providing a method to derive the free
energy difference between the initial and final states. The reversible
work $\Delta G_{X_{\rm min}}^{X_{\rm max}}$ gets contributions from
pulling the bead and stretching the handles and the ssDNA (see
\ref{apc}). In particular, we have defined $\Delta G_1(f_{\rm
  max})$ in \eref{wrevX3} as that part of the reversible work that gets
contributions only from unfolding and stretching the hairpin,
\be
\Delta G_1(f_{\rm max})=\Delta G_{X_{\rm min}}^{X_{\rm max}}-\frac{1}{2k_{\rm eff}}\left[(f_{\rm
    max})^{2}-(f_{\rm min})^2\right]\,.
\label{dg1ft}
\ee
In what follows we subtract for each molecule from the value of the
work $W$ the term $\frac{1}{2k_{\rm eff}}\left[(f_{\rm
    max})^{2}-(f_{\rm min})^2\right]$ to directly estimate the
contribution to the free energy of the hairpin, $\Delta G_1(f_{\rm
  max})$. The value of $\Delta G_1(f_{\rm max})$ can be derived by
looking at the crossing of the stretching and releasing work
distributions. Although stretching and releasing work distributions
for the same molecule taken at different speeds cross at a common
value $\Delta G_1(f_{\rm max})$ (data not shown), work histograms
mostly change from molecule to molecule revealing some variability in
our estimates for $\Delta G_1(f_{\rm max})$. This is probably due to
the fact that we are using very short tethers as handles which, due to
their large rigidity and depending on the angle formed by the tether
connecting the two beads, introduce a high variability to the free
energy correction $\frac{1}{2k_{\rm eff}}\left[(f_{\rm
    max})^{2}-(f_{\rm min})^2\right]$.

\begin{figure}
\begin{center}
\vspace{0.9cm}
\includegraphics[width=7cm,angle=0]{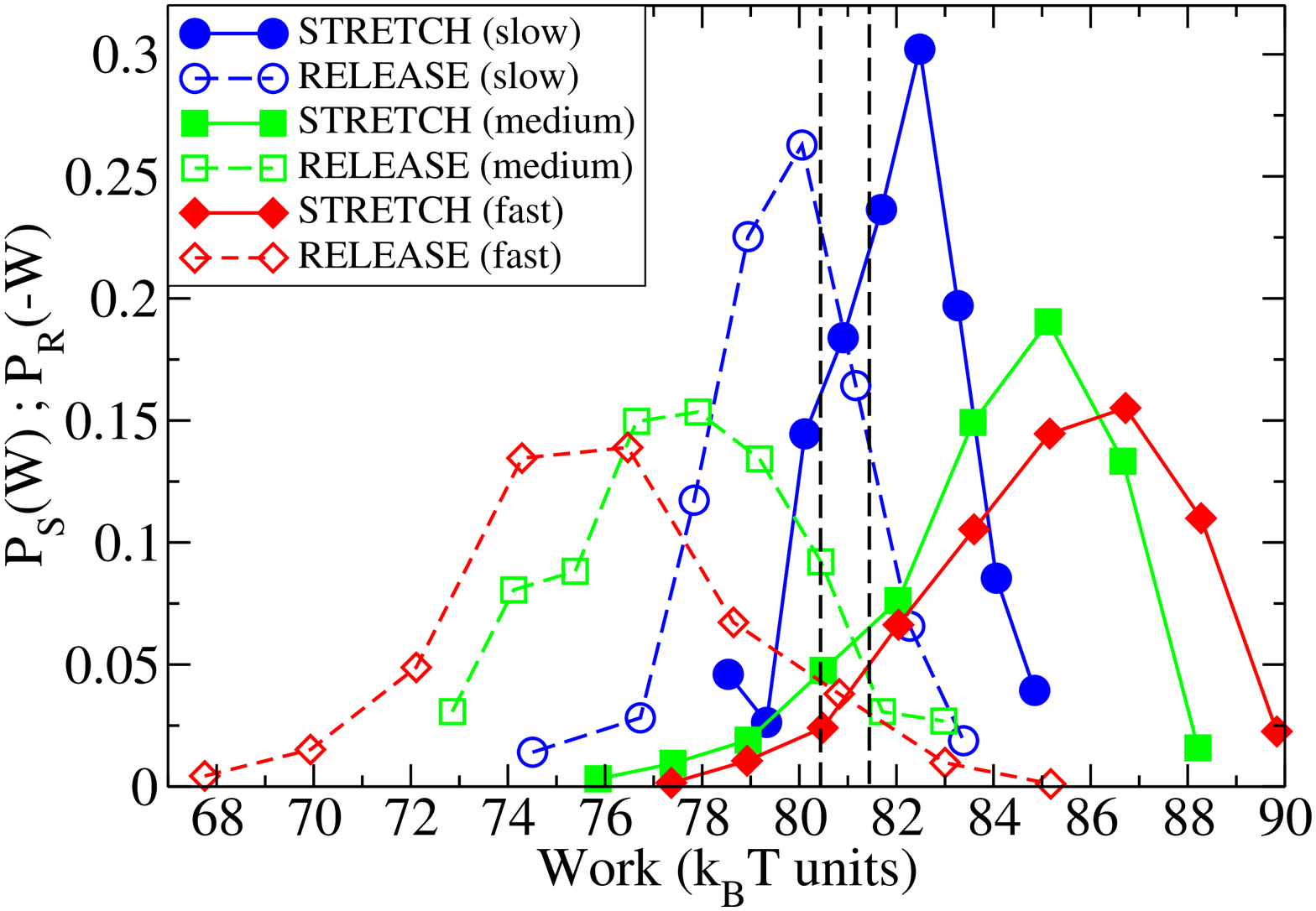}\hspace{.5cm}\includegraphics[width=7cm,angle=0]{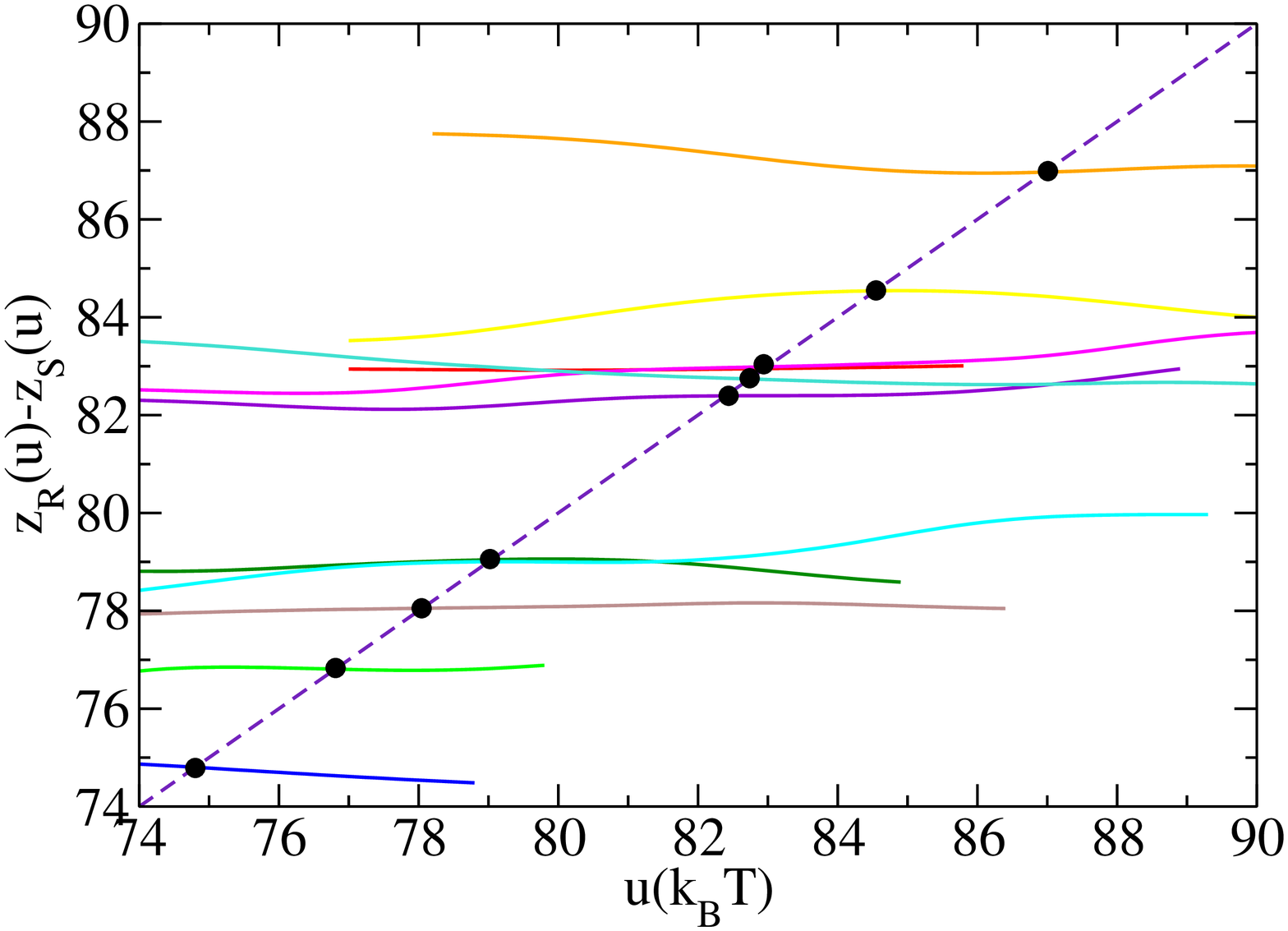}
\caption{ (Left) Typical work distributions for three molecules at three
  different loading rates: 1 pN/s (slow, blue), 4.88 pN/s (medium, green),
  14.9 pN/s (fast, red). Work values have been shifted in such a way that the crossing
  between the stretching and releasing work distributions
  is observed at the average value $\Delta G_1^{\rm FR}=80.94 k\sub{B}T$ (see
  \tref{table3}). The vertical lines show the range of experimental
  error estimated for the value of $\Delta G_1^{\rm FR}$. (Right) Bennett
aceptance ratio method to extract the value of $\Delta G_1(f_{\rm max})$
for all 11 molecules (shown as black dots).}
\label{fig9}
\end{center}
\end{figure}

In \fref{fig9} (left panel) we show work distributions for three different molecules pulled at different speeds. To
better show the systematic dependence of work distributions on the
pulling speed, the histograms shown in \fref{fig9} (left panel) have been
shifted to make the crossing point (between stretching and releasing
distributions) coincide with the value of $\Delta G_1(f_{\rm max})$
averaged over all molecules (see below and \tref{table3}).

In order to validate the fluctuation relation \eref{crooksfr} and extract
the value of $\Delta G_1(f_{\rm max})$ for each molecule we have analyzed data in two
ways:

\subsubsection*{ Bennet acceptance ratio method.} The details of this method have been
described elsewhere \cite{Ritort08,ColRitJarSmiTinBus05}. In a nutshell, Bennett's method consists in defining
two functions,
\numparts
\be
z\sub{S}(u)&=\left\langle  g_{u}(W)\exp\left(-\frac{W}{k\sub{B}T}\right)\right\rangle\sub{R} \,,\label{zs}\\
z\sub{R}(u)&=\log\left(\langle g_{u}(W)\rangle\sub{S}\right)\,,
\label{zr}
\ee
\endnumparts
where $ g_{u}(W)$ is an arbitrary real function that depends on a
parameter $u$ and the average $\langle\dots\rangle\sub{S(R)}$ is taken over the
stretching (releasing) process. From \eref{crooksfr}, it can be proven that 
\be
z\sub{R}(u)-z\sub{S}(u)=\frac{\Delta G_{X_{\rm min}}^{X_{\rm max}}}{k\sub{B}T} \,,
\label{zrzs}
\ee
showing that the difference between the two functions is constant over
$u$ and equal to the reversible work. It has been shown by Bennett
\cite{Bennett76} that the optimal (i.e.\ minimal variance) estimate of $\Delta
G_{X_{\rm min}}^{X_{\rm max}}$ is given by 
\be
g_{\mu}(W)=\frac{1}{1+\frac{N\sub{S}}{N\sub{R}}\exp\left(\frac{W-u}{k\sub{B}T}\right)} \,,
\label{bio:accep:6}
\ee
with $u=\Delta G_{X_{\rm min}}^{X_{\rm max}}$. $N\sub{S(R)}$ stands for the
number of pulls along the stretching (releasing) process. The same result has been obtained by Pande and
coworkers by using maximum likelihood methods \cite{ShiBaiHooPan03}. In
\fref{fig9} (right panel) we show the test applied to work data
for all molecules. The experimental data for $z\sub{R}(u)-z\sub{S}(u)$ is
approximately constant for each molecule over a wide range of $u$. The
best estimate for $\Delta G_1(f_{\rm max})$ in \eref{dg1ft} is obtained
by looking at the intersection of the experimental data with the line
$z\sub{R}(\mu)-z\sub{S}(u)=u/k\sub{B}T$ (black dots in the figure). As we can see
the fluctuation relation is validated for each molecule. However there
is a strong variability from molecule to molecule for the values $\Delta G_1(f_{\rm max})$.

\subsubsection*{ Direct representation of the probability ratio.} 
\begin{figure}
\begin{center}
\vspace{.9cm}
\includegraphics[width=7.cm,angle=0]{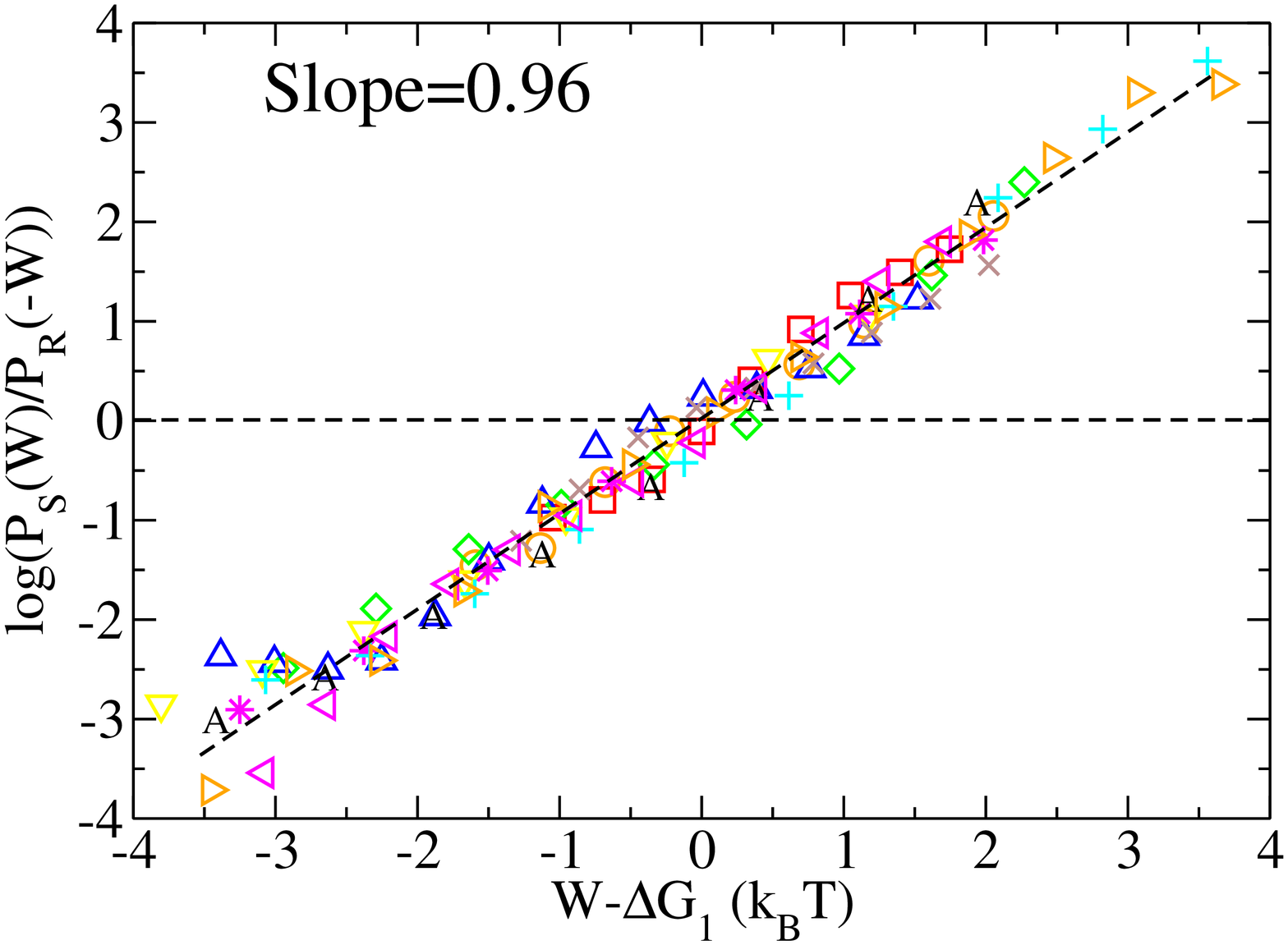}\hspace{0.3cm}\includegraphics[width=7.5cm,angle=0]{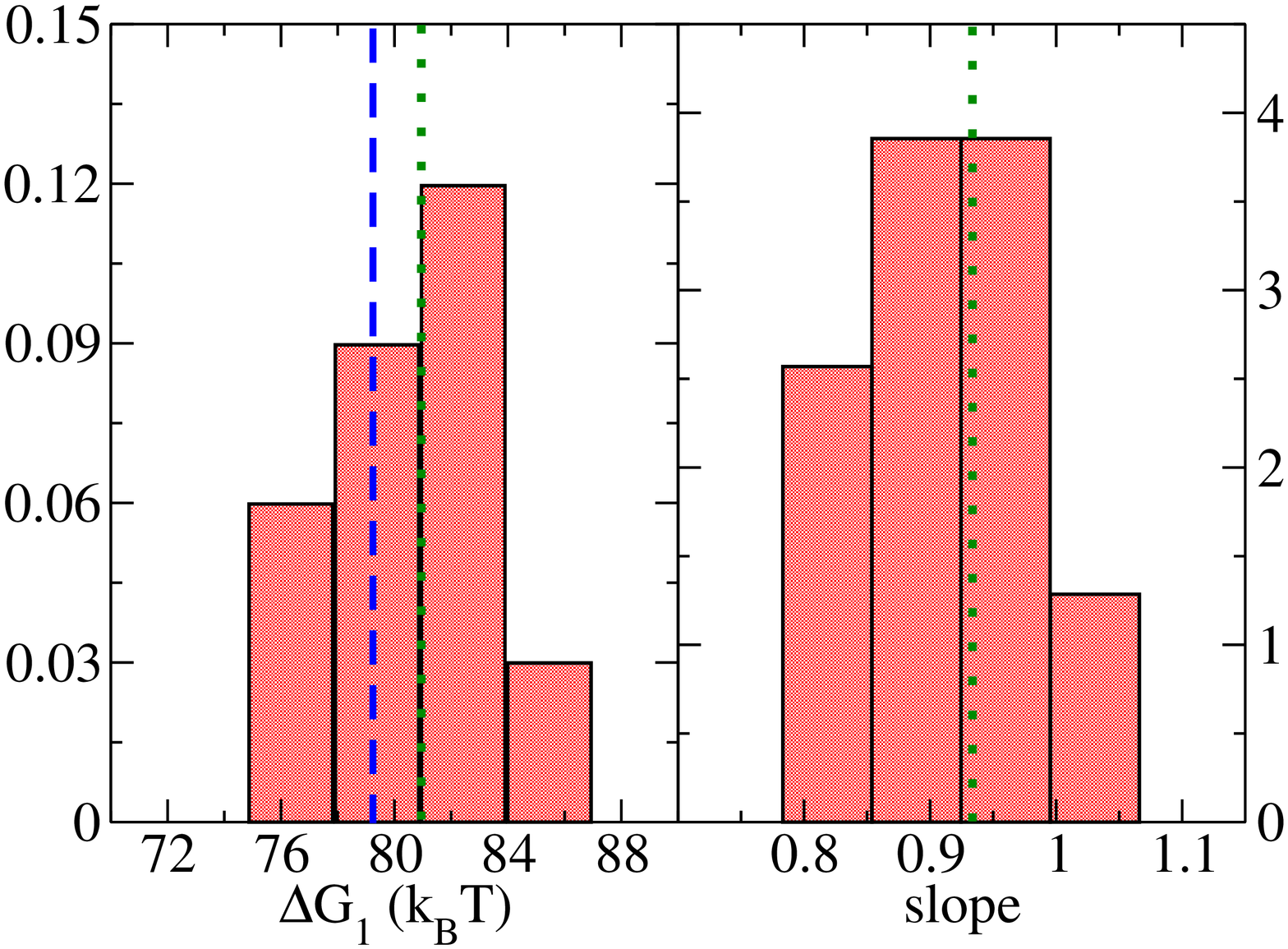}
\caption{ (Left) Experimental verification of the Crooks fluctuation
  relation \eref{crooksfr}. For each molecule, work data has been
  shifted to cross the horizontal axis (dashed line) at $W=\Delta
  G_1(f_{\rm max})$ (corresponding to the black dots shown in
  \fref{fig9}, right panel). Different symbols correspond to the 11
  molecules. The continuous line is a weighted linear fit to all data
  that has slope equal to 0.96. (Right) Histograms of the values of
  $\Delta G_1(f_{\rm max})$ and the slopes corresponding to data shown
  in the left panel. Statistics collected over 11 molecules. The
  vertical dotted lines show the mean of the histogram whereas the
  dashed line in the histogram of $\Delta G_1(f_{\rm max})$ indicates
  the average value of $\Delta G_1$ that has been obtained from
  kinetics (\sref{kineticpar} and \tref{table1}).}
\label{fig10}
\end{center}
\end{figure}

The validity of the fluctuation relation \eref{crooksfr} is again
observed in \fref{fig10} (left panel) where we have plot the ratio in the l.h.s.\
of \eref{crooksfr} in logarithmic scale versus the work $W$. Like we did
with the work histograms shown in \fref{fig9} (left panel), work values have
been shifted for the estimate $\Delta G_1(f_{\rm max})$ to match the
the average value $\Delta G_1^{\rm FR}=80.94\ k\sub{B}T$ over all molecules (\tref{table3}).
Finally, in \fref{fig10} (right panel) we plot the histogram of values for $\Delta
G_1(f_{\rm max})$ obtained for different molecules. As we already saw
for kinetics, the value of $\Delta G_1(f_{\rm max})$ changes from molecule to molecule, yet the average value $\Delta G_1^{\rm
  FR}=80.94\ k\sub{B}T$  (\tref{table3}) is compatible with the
estimate obtained from kinetics $\Delta
G_1=78.73$ (\sref{kineticpar} and \tref{table1}). In
\fref{fig10} (right panel) we also show the histogram of the
different slopes of the fluctuation relation shown in the left
panel. The fluctuation relation is reasonably well satisfied by the
experimental data with an
average slope of 0.93. However, if we weight the different slopes
according to the number of pulls for each molecule we obtain 0.96, which is closer to the expected
value of 1.

\section{Derivation of the value of $\Delta G_0$}
\label{dg0}

In the preceding sections we showed ways of extracting the values of
$\Delta G_1$ and $\Delta G_{X_{\rm min}}^{X_{\rm max}}$ by using
rupture force kinetics or the fluctuation relation. Now we face the problem
of extracting the value of the free energy of formation of the hairpin
at zero force, $\Delta G_0$, using both methods. As we saw in
\eref{critf} and \eref{wrevX4}, and in order to extract $\Delta G_0$, we must subtract from the total energy
$\Delta G_1$ and $\Delta G_{X_{\rm min}}^{X_{\rm max}}$ the stretching
contribution to the free energy,
\be
\fl G_{\rm ssDNA}(0\to x^{l_{N}};f)=\int_{0}^{x^{l_{N}}}F_{\rm ssDNA}^{l_N}(y)\rmd y\,,\quad f=F_{\rm ssDNA}^{l_N}(x_{N}) \,,
\label{dg01}
\ee
with $f=f\sub{c}$ in \eref{critf} and $f=f_{\rm max}$ in
\eref{wrevX4}. $l_N=(2N+L)d$ denotes the full contour length of the hairpin
(\sref{free_ene}). For the
elastic response of the ssDNA, $F_{\rm ssDNA}^{l}(x)$, we use the freely-jointed chain model
\cite{SmiCuiBus95,HugForSmiBusRit08},
\be
x(f)=l\left(1+\frac{f}{Y}\right)\left[\coth\left(\frac{fb}{k\sub{B}T}\right)-\frac{k\sub{B}T}{fb} \right] \,,
\label{dg02}
\ee
with $l$ the contour length of the ssDNA, $b$ the Kuhn length and $Y$
the Young modulus. For our temperature and salt conditions we take
$b=1.43$ nm, $Y=812$ pN \cite{SmiCuiBus95,HugForSmiBusRit08}. 

We must stress that in order to extract $\Delta G_0$ from either rupture
force kinetics data or by using the fluctuation relation, it is necessary to take into
account the fact that the force is not controlled and adopt expressions
derived in the appropriate experimental mixed ensemble. In what follows
we consider both cases.

\subsubsection*{\bf Deriving $\Delta G_0$ from rupture force kinetics.} From the value obtained for
  $\Delta G_1$ from rupture force kinetics (\tref{table1}), we now
  adopt the expression \eref{critf} with $\Delta G_1(f\sub{c})=\Delta G_1$
  where $\Delta G_1$ has been measured from kinetics as explained in
  \sref{simplified}. However, using that expression would lead to
  incorrect results for $\Delta G_0$. The reason is that in our
  experiments the force is not controlled, as we only control the position
  of the trap. As we have shown in \ref{apc}, a contraction
  in molecular extension (induced by the finite diameter of the hairpin)
  shifts the free energy of the fully unfolded state, relative to any
  partially unfolded intermediate state, by an amount equal to
  $-f_\mathrm {c} d_0$. Therefore,
\be
\fl \Delta G_0=\Delta G_1-G_{{\rm ssDNA}}(0\to x_{N};f\sub{c})+f\sub{c} d_0\,, \qquad
f \sub{c}=F_{\rm ssDNA}^{l_N}(x_{N})\,.
\label{dg03}
\ee
Although the values obtained for $\Delta G_0$ show the same dispersion as
we saw for $\Delta G_1$ in \sref{kineticpar} (see \fref{fig8}),
the average value of $\Delta G_0$ is not far from the Mfold \cite{Zuker2003}
predicted value (\fref{fig11} and
\tref{table2}).

\subsubsection*{\bf Deriving $\Delta G_0$ from the fluctuation relation.} 
\begin{figure}
\begin{center}
\vspace{.9cm}
\includegraphics[width=8cm,angle=0]{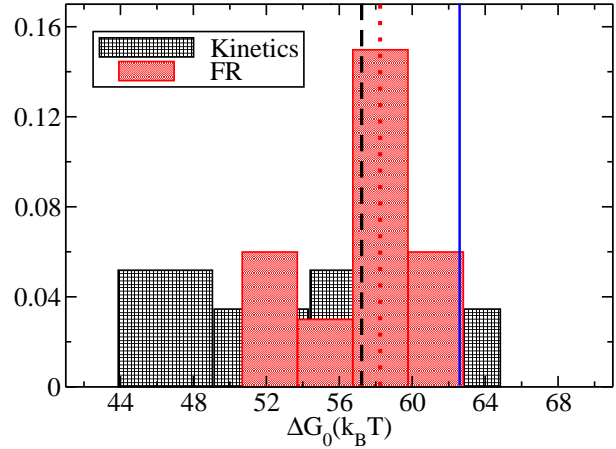}
\caption{  Histograms of the values of
  $\Delta G_0$ obtained from kinetics (\tref{table2}) and the
  fluctuation relation (\tref{table3}). The vertical dashed line indicates
  the best estimate for $\Delta G_0$ obtained from kinetics (black),
  the vertical dotted line indicates the best estimate obtained
  from the fluctuation relation (red) and the vertical continuous line
  the value predicted by Mfold \cite{Zuker2003} (blue) at the
  experimental conditions ($23^{\circ}C$ and 1M NaCl).}
\label{fig11}
\end{center}
\end{figure}

We now
use \eref{g1mixedb},
\be
\Delta G_0=\Delta G_1(f_{\rm max})-G_{\rm ssDNA}(0\to
x^{l_{N}};f_{\rm max})+f_{\rm max}d_0 \,,
\label{dg04}
\ee
with $f_{\rm max}$ the maximum force along the force cycles. We have
determined the value of $\Delta G_0$ for each molecule using the
previously determined values for $\Delta G_1(f_{\rm max})$ in \sref{crooks}. As we saw in
\fref{fig8}, there is also some variability for the values of $\Delta G_0$
obtained for different molecules using this method. However, the average of the different
values is not far from the value expected from Mfold \cite{Zuker2003}
(\fref{fig11} and \tref{table2}).

\begin{table}
 \caption{Mean and standard deviations of free energy parameters obtained
from the kinetic measurements. Averages are taken over 11 different
molecules that have been pulled at different
pulling speeds:  18.5 nm/s or 1 pN/s (1 molecule, 193 cycles), 36.5 nm/s or 1.95 pN/s (2
molecules, 160 cycles), 86.2 nm/s or 4.88 pN/s (3 molecules, 570 cycles), 156 nm/s or 8.1 pN/s (2
molecules, 725 cycles), 274.3 nm/s or 14.9 pN/s (3 molecules, 1580
cycles). \label{table2}}
\begin{indented}
\item[]\begin{tabular}{cccc}
\br
$f\sub{c}$ (pN) & $x_{\rm ssDNA}$ (nm) & $d_0$ (nm) & $(\Delta G_{\rm ssDNA})^{\rm
    kin}$ ($k\sub{B}T$) \\
\mr
17.91 & 23.3 & 2.0 & 30.24 \\
\br
$(\Delta G_1)^{\rm kin}$ ($k\sub{B}T$) & $(\Delta G_0)^{\rm
  kin}$ ($k\sub{B}T$) & $(\Delta G_0)^{\rm kin}$ (kcal/mol) & Mfold
(kcal/mol)\\
\mr
78.7(2.4) & 57.2(2.4) & 33.7(1.5) & 36.81\\
\br
\end{tabular}
\end{indented}
\end{table}
\begin{table}
 \caption{Mean and standard deviations of free energy parameters obtained
from the Crooks fluctuation relation. Averages are taken over 11 different
   molecules that have been pulled at different pulling speeds: 18.5 nm/s
   or 1 pN/s (1 molecule, 193 cycles), 36.5 nm/s or 1.95 pN/s (2
   molecules, 160 cycles), 86.2 nm/s or 4.88 pN/s (3 molecules, 570 cycles),
   156 nm/s or 8.1 pN/s (2 molecules, 725 cycles), 274.3 nm/s or 14.9 pN/s (3
   molecules, 1580 cycles). The value for $\Delta (f^2/2k_{\rm eff})$
   depends on the maximum and minimum force that varies for each
   molecule. Therefore we just show the average of this number without
   giving the error. \label{table3}}
\begin{indented}
\item[]\begin{tabular}{cccc}
\br
$x_{\rm ssDNA}$ (nm) & $d_0$ (nm) & $(\Delta G_{\rm ssDNA})$ ($k\sub{B}T$) & $\Delta (f^2/2k_{\rm eff})$ ($k\sub{B}T$)\\
\mr
23.70(03) & 2.0 & 32.40(04) & 341.35\\
\br
$(\Delta G_1)^{\rm FR}$ ($k\sub{B}T$)&$(\Delta G_0)^{\rm FR}$ ($k\sub{B}T$) & $(\Delta G_0)^{\rm FR}$ (kcal/mol) & Mfold (kcal/mol)\\
\br
80.9(1.0) & 58.2(1.0) &  34.2(6) & 36.81\\
\br
 \end{tabular}
\end{indented}
\end{table}

\section{Conclusions}

In this work we have investigated the mechanical unfolding of DNA hairpins
using optical tweezers. We have tested the validity of Kramers--Bell
theory for force dependent folding/unfolding kinetic rates by comparing
theoretical predictions to single molecule pulling experiments on DNA
hairpins. We introduced the concept of free energy landscape for generic
nucleic acid (DNA and RNA) molecules and derived simplified expressions
for the Kramers--Bell kinetic rates. To validate the theoretical
predictions we have carried out experiments on a specifically designed
DNA sequence that displays cooperative two-state behaviour. According to
theory, this sequence has a free energy landscape characterized by two
states (folded and unfolded) that are separated by a single barrier and
a transition state located in a position along the molecular sequence
that is independent on the applied force. By doing rupture force
measurements we are capable of predicting the main parameters that
characterize the free energy landscape of the hairpin, such as: the
distances of the folded and unfolded states to the transition state
($x^{\rm F},x^{\rm UF}$), the free energy difference between both states
($\Delta G_1$) and the coexistence rate ($k\sub{c}$). By measuring the
mechanical work and using the Crooks fluctuation relation we can also
extract the reversible work in our experiments. Both type of
measurements (rupture force kinetics and the fluctuation relation) yield
values for the free energy of formation of the hairpin at zero force,
$\Delta G_0$, that are compatible with each other and with the
calorimetric based prediction by Mfold. We remark the following results:

\begin{itemize}

\item{\bf Validity of the Kramers--Bell simplified rates.} The low
  value of the released molecular extension $x\sub{m}$ obtained in
  kinetic studies suggests that the Kramers--Bell model described in
  \sref{free} is an over-simplification of the true folding/unfolding
  kinetics of the hairpin. It provides estimates for the kinetic
  parameters that fit the expected values obtained from theory within
  $15\%$, but cannot do better. Another possible explanation for such
  lower values of $x\sub{m}$ is molecular fraying at the beginning of
  the hairpin stem that reduces the molecular length and free energy
  of the hairpin. Given the complexity of the molecules we are
  investigating, a $15\%$ agreement between theory and experiment can
  be considered reasonably good. Yet it would be desiderable to
  explore different sequences and molecular constructions and develop
  models that can improve the agreement between theory and experiments.

\item{\bf Free energy recovery.} A noticeable result from our
  experiments is the strong variability observed for the parameters
  extracted from rupture force kinetic studies for different molecules
  pulled at different speeds (\tref{table1} and
  \fref{fig8}). Variability between parameters is an inherent aspect
  of single molecule experiments. Concerning the free energy of the
  hairpin $\Delta G_0$, and as shown in \fref{fig11}, the fluctuation
  relation by Crooks is compatible with the estimate obtained from
  rupture force kinetics. Both values are 2-3kcal/mol below the value
  predicted by Mfold leading to a discrepancy between
  5-10\%. Yet, the fluctuation relation provides less spread values
  and more reliable final estimates for $\Delta G_0$. Many sources of
  error can account for such discrepancy: limitations of the
  nearest-neighbour base pair model used to describe the
  thermodynamics of unzipping as well as uncertainty in the Mfold
  nearest neighbour free energies; the innacurate knowledge of the
  elastic properties of the released ssDNA; the innacurate
  determination of the diameter of the hairpin; molecular fraying and;
  force calibration and other experimental errors.

\item{\bf The diameter of the hairpin $d_0$ and the elastic properties
  of the ssDNA.} The value of $d_0$ ($\simeq 2$ nm) taken from
  structural studies of the double helix, and the polymer model used
  to model the elastic response of the ssDNA, are particularly
  important for correctly estimating the value of $\Delta G_0$. For
  example, a 20\% error in the value of $d_0$ introduces a 1 kcal/mol
  error in $\Delta G_0$ ($\simeq 1.6\ k\sub{B}T$ at room
  temperature). How reliable is our value for $d_0$? How accurate are
  the parameters of the freely jointed chain (Kuhn length, Young
  modulus and interbase distance) to describe the elastic properties
  of the ssDNA released by the hairpin? It would be much interesting
  to carry out similar detailed investigations in other DNA
  sequences. In this way we could check whether the values we adopted for
  these parameters are generically accurate for arbitrary hairpin
  sequences (or instead they depend on the DNA sequence).
  
\end{itemize}
Two-state models are very useful to address questions related to
thermodynamics and kinetics of force induced transitions. They guide us
in validating and detecting limitations of current theories and models
describing the folding of nucleic acids and proteins. Two-state models are also useful
to investigate issues related to the irreversibility and dissipation of
nonequilibrium small systems as reported in our companion paper
\cite{ManMosForHugRit08}. Finally, the current investigation is a first
step to approach the force-induced folding/unfolding kinetics of more complex molecular structures.

\ack

We acknowledge financial support from grants FIS2007-61433, NAN2004-9348, SGR05-00688.

\appendix
\setcounter{section}{0}

\section{Free energy landscape in the mixed ensemble \label{apa}} 

In \sref{free} we assumed that the experimentally
controlled variable is the external force. However, this is far from true in single molecule
experiments with optical tweezers or atomic force microscopy, where the force is a fluctuating
variable\footnote{Exceptions are magnetic tweezers \cite{GosCro02} or specifically
  designed tweezers setups with zero stiffness regions \cite{GreWooAbbBlo05}. Force feedback
  systems are not ideal constant force systems as they introduce other
  sort of noise effects due to the limited feedback frequency (typically
  around 1 kHz).}. Is it possible to define and compute free energy
landscapes beyond the force ensemble worked out in \sref{free}? Here
we show how to extend the concept of free energy landscape to the mixed
ensemble case relevant for our pulling experiments. The presentation
here is summarized, the interested reader will find details in
\cite{ManRit05}.
\begin{figure}
\begin{center}
\includegraphics[width=12cm,angle=0]{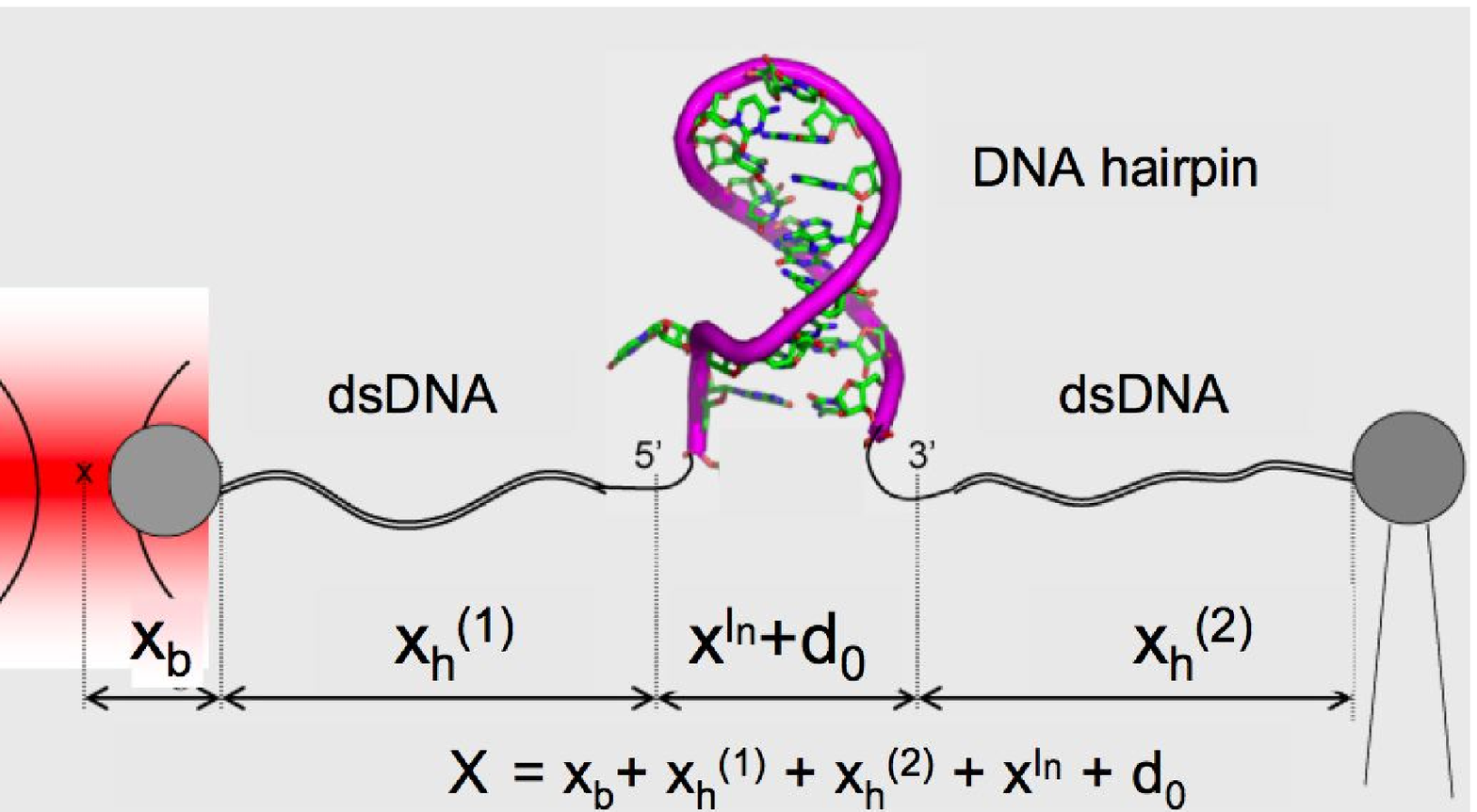}
\caption{ Schematics of the mixed ensemble. The total distance $X$ is
  expressed as the sum of the different extensions $x\sub{b},x\sub{h},x_n$. $x\sub{h}=x\sub{h}^{(1)}+x\sub{h}^{(2)}$  is the total length of the handles, while the extension $x_n$ has been defined in \eref{xn(f)}. The hairpin diameter $d_0$ is taken to be equal to 2 nm and must be included in $x_n$ for all  configurations of the hairpin with the exception of the fully unfolded one.}
\label{fig12}
\end{center}
\end{figure}

A schematic representation of the relevant experimental setup
corresponding to the mixed ensemble is shown in \fref{fig12}. Let
$X,x_n,x\sub{b},x\sub{h}$ denote the trap--pipette distance, the molecular
extension of the hairpin \eref{xn(f)}, the bead position in the trap and the handles
total extension, respectively. These satisfy $X=x\sub{b}+x\sub{h}+x_n$. The force is given by
$f=k\sub{b}x\sub{b}$, where $k\sub{b}$ is the stiffness of the trap. In the mixed
ensemble, the molecular extension of the hairpin plus handles ($\ell_n\equiv x_n+x\sub{h}$)
and the force ($f$) are fluctuating variables and only $X$ is externally
controlled. Similarly to what we did in \sref{free}, let $G(\ell_n,X)$
denote the free energy necessary to break the first $n$ base pairs
along the hairpin (starting from the beginning of the fork), thus generating a molecular
extension (bead-to-bead) $\ell_n$ when the trap--pipette distance is equal to $X$. We can write
\be
G(\ell_n,X)=G(\ell_n,0)+G_{\rm stretch}(0\to \ell_n;X)\,,
\label{gxX}
\ee
where $G_{\rm stretch}(0\to\ell_n; X)$ gets contributions from the bead, the
handles and the ssDNA released by the hairpin:
\be
\fl G_{\rm stretch}(0\to\ell_n;X)=G_{\rm b}(0\to x\sub{b};X)+G_{\rm h}(0\to x\sub{h};X)+G_{\rm ssDNA}(0\to x^{l_n}; X)\,.
\label{work_stretchX}
\ee
The initial condition $X=0$ in the term $G(\ell_n,0)$ in \eref{gxX} must be
understood as that position of the trap where all elements are fully
relaxed and subject
to zero tension.  The different terms in \eref{work_stretchX} are given by 
\be
G_{\rm b}(0\to x\sub{b};X)&=\int_{0}^{x\sub{b}}F_{\rm b}(y)\rmd y \,, \label{gbead}\\
G_{\rm h}(0\to x\sub{h};X)&=\int_{0}^{x\sub{h}}F_{\rm h}(y)\rmd y \,, \label{ghandle}\\
G_{\rm ssDNA}(x_0\to x^{l_n};X)&=\int_{0}^{x^{l_n}} F_{\rm ssDNA}^{l_n}(y)\rmd y-F_{\rm ssDNA}^{l_N}(x_N)(1-\delta_{n,N}) \,, \label{gssDNA}
\ee
where $F_{\rm b}(y)=k\sub{b}y$ is the elastic response of the bead,
$F_{\rm h}(y)$ stands for the equilibrium force-extension curve for the handles and $F_{\rm ssDNA}^{l_n}(y)$ is the equilibrium
force-extension curve for the ssDNA of contour length $l_n$. The last term in \eref{gssDNA} accounts for the shortening of the molecular extension equal to the diameter of the hairpin that occurs when the last base pair of the hairpin unzips. The contour
length $l_n$ in \eref{gssDNA} satisfies the mechanical equilibrium conditions, 
\be
\fl f=F_{\rm b}(x\sub{b})=F_{\rm h}(x\sub{h})=F_{\rm ssDNA}^{l_n}(x^{l_n})\,, \qquad 
X=x\sub{b}+x\sub{h}+x^{l_n}+d_0(1-\delta_{n,N}) \,,
\label{equacioX}
\ee
defining also the Lagrange multiplier $f$ (corresponding to the average
of the in\-stan\-ta\-ne\-ous and fluctuating force acting on each element). Note
that for a given pair $(\ell_n,X)$ we have 3 unknowns ($x\sub{b},x\sub{h},x_n$). The three
independent equations in \eref{equacioX} fully determine the system so we
can exactly compute $G(\ell_n,X)$ as a function of $\ell_n$ for a given value of
$X$. For that we must know the elastic response of the different
elements: $F_{\rm b},F_{\rm h},F_{\rm ssDNA}^l$.

\section{Dependence of $B_1(f),\Delta G_1(f)$ across the transition \label{apb}}

Here we show that terms of the type $G_{\rm ssDNA}(0\to x^{l_{\rm F}};f)$ entering in
\eref{g1a}, \eref{g2a} vary with $f$ much less than the corresponding product
$fx^{\rm F}$ does. In other words,
\be
\frac{\partial G_{\rm ssDNA}(0\to x^{l_{\rm F}};f)}{\partial f}\ll x^{\rm
  F}+f\frac{\partial x^{\rm F}}{\partial f} \,,
\label{apb1}
\ee
where we assume that $x^{\rm F}$ generally depends on $f$ (see \sref{free_ene}). 
We start from \eref{work_stretch}, \eref{equaciof} and write
\be
\frac{\partial}{\partial f} G_{\rm ssDNA}(0\to x^{l_{\rm F}};f)=\int_0^{x^{l_{\rm F}}}\frac{\partial F_{\rm ssDNA}^{l_{\rm F}}(y)}{\partial f}\rmd y+f\frac{\partial x^{\rm F}}{\partial f} \,,
\label{apb2}
\ee
where $F_{\rm ssDNA}^{l\sub{F}}(y)$ in the integrand depends on $f$ through the
dependence of the contour length as given in \eref{equaciof}. The elastic
response of biopolymers (e.g. ssDNA) satisfy either the worm-like chain
model or the freely jointed chain model. In both cases we have that the
force is sole function of the extension divided by the contour length,
\be
F_{\rm ssDNA}^l(y)\equiv \hat{F}_{\rm ssDNA}(y/l)\,.
\label{apb3}
\ee
It is also reasonable to assume that the relative change in the contour length $l_F$ of the released
ssDNA is at most equal to the relative change of the distance $x^{\rm F}$,
\be
\frac{1}{l\sub{F}}\frac{\partial l\sub{F}}{\partial f}\le \frac{1}{x^{\rm F}}\frac{\partial x^{\rm F}(f)}{\partial f}\,.
\label{apb4}
\ee
From \eref{apb3}, \eref{apb4} we can write
\be
\frac{\partial F_{\rm ssDNA}^{l\sub{F}}(y)}{\partial f}=\frac{\partial F_{\rm
    ssDNA}^{l\sub{F}}(y)}{\partial l\sub{F}}\frac{\partial l\sub{F}}{\partial f}\le -\frac{y}{x^{\rm F}}\frac{\partial F_{\rm
    ssDNA}^{l\sub{F}}(y)}{\partial y}\frac{\partial x^{\rm F}}{\partial f}\,.
\label{apb5}
\ee
Inserting this expression in the integrand in \eref{apb2} and doing an
integration by parts we obtain
\be
\int_0^{x^{\rm F}}\frac{\partial F_{\rm ssDNA}^l(y)}{\partial
  f}\rmd y\le -\frac{1}{x^{\rm F}}\frac{\partial x^{\rm F}}{\partial
  f}\left(fx^{\rm F} -\int_0^{x^{\rm F}}F_{\rm ssDNA}^l(y)\rmd y\right)\,.
\label{apb6}
\ee
Plugging this expression into \eref{apb2} we get
\be
\fl \frac{\partial G_{\rm ssDNA}(0\to x^{\rm F};f)}{\partial f}\le \frac{1}{x^{\rm F}}\frac{\partial x^{\rm F}}{\partial
  f}\int_0^{x^{\rm F}}F_{\rm ssDNA}^l(y)\rmd y\le f\frac{\partial x^{\rm
    F}}{\partial f}\ll x^{\rm F} +f\frac{\partial x^{\rm F}}{\partial f} \,,
\label{apb7}
\ee
where we introduced the upper bound $F_{\rm ssDNA}^l(y)\le f$ in the
integrand. The last inequality is generally valid for short hairpins
that are characterized by a rigid ssDNA where, according to
\eref{apb3},
\be
k_{\rm ssDNA}=\frac{\partial f}{\partial x^{\rm F}}\gg \frac{f}{x^{\rm F}}\,.
\label{apb8}
\ee
Therefore we proved \eref{apb1}. The same demonstration applies for the
term  $G_{\rm ssDNA}(0\to x^{l_{N}};f)$, proving that we can approximate \eref{rate_2sa}, \eref{rate_2sb} by the simplified rates \eref{e7}. 

\section{Reversible work in the mixed ensemble \label{apc}} 

Here we show which terms contribute to the experimentally measured
reversible work $\Delta G_{X_{\rm min}}^{X_{\rm max}}$ appearing in the Crooks
fluctuation relation \eref{crooksfr}. From the definition of $G(x,X)$ in \ref{apa}, we
can write 
\be
\Delta G_{X_{\rm min}}^{X_{\rm max}}=G(x_{N}^{\rm max},X_{\rm max})-G(x_0,X_{\rm min}) \,,
\ee
where $x_{N}^{\rm max}$ is the molecular extension of the fully unfolded hairpin at
$X=X_{\rm max}$.  From
\eref{gxX},\eref{work_stretchX},\eref{gbead},\eref{ghandle} and \eref{gssDNA} we obtain
\be
\Delta G_{X_{\rm min}}^{X_{\rm max}}=\int_{x\sub{b}^{\rm min}}^{x\sub{b}^{\rm
    max}}F_{\rm b}(y)\rmd y+\int_{x\sub{h}^{\rm min}}^{x\sub{h}^{\rm max}}F_{\rm
  h}(y)\rmd y+\int_{0}^{x_{N}^{\rm max}}F_{\rm
  ssDNA}^l(y)\rmd y\nonumber\\
-f_{\rm max}d_0+\Delta G_0 \label{wrevx}
\ee
where $\Delta G_0$ was defined in \eref{dgf} and where we have introduced
the equilibrium force at the end of the stretching cycle at $X_{\rm
  max}$, $f_{\rm max}=F_{\rm ssDNA}^{l_N}(x_{N}^{\rm max})$. Note
that the correction term $f_{\rm max}d_0$ induced by the finite diameter
of the hairpin has a different sign in the force ensemble
\eref{dgfold}. The reason is that, in the force ensemble, an increase in
the total distance $X$ of the system (at fixed force $f$) decreases the
total free energy of the system. However, in the mixed ensemble, the
same increase in the total distance $X$ (followed by the corresponding
drop in the force $f$) increases the total free energy of the system.

We now introduce the force-dependent rigidity of the handles,
\be
k\sub{h}(f)=\left( \frac{\partial F_{\rm h}(y)}{\partial y} \right)_{f=F_{\rm h}(y)}\,,
\label{rigidh}
\ee
and use \eref{equacioX} to change all integration variables in \eref{wrevx}
to the Lagrange multiplier (force) $f$. We finally get
\be
\fl \Delta G_{X_{\rm min}}^{X_{\rm max}}=\int_{f_{\rm min}}^{f_{\rm
    max}}\frac{f\rmd f}{k\sub{b}}+\int_{f_{\rm min}}^{f_{\rm max}}\frac{f\rmd f}{k\sub{h}(f)}+\int_{0}^{x_{N}^{\rm max}}F_{\rm
  ssDNA}^l(y)\rmd y-f_{\rm max}d_0+\Delta G_0\,, \label{wrevX2}
\ee
where we have written $f_{\rm max}=F_{\rm  ssDNA}^{l_N}(x_{N}^{\rm max})$.
If the handles are very rigid compared to the rigidity of the trap then $k\sub{h}\gg k\sub{b}$ and therefore the sum of
the two integrands in the rhs of \eref{wrevX2} can be approximated by 
\be
\frac{1}{k_{\rm eff}}=\frac{1}{k\sub{b}}+\frac{1}{k\sub{h}(f)}\,,
\label{keff}
\ee
where $k_{\rm eff}$ is taken as constant in the interval $[f_{\rm
    min},f_{\rm max}]$. The condition $k\sub{h}\gg k\sub{b}$ is reasonably well
satisfied in our experiments where we use 29-bp handles. Therefore, by
substituting \eref{keff} and defining
\be
\Delta G_1(f_{\rm max})=\Delta G_{X_{\rm min}}^{X_{\rm
    max}}-\int_{f_{\rm min}}^{f_{\rm  max}}\frac{f\rmd f}{k\sub{b}}-\int_{f_{\rm
    min}}^{f_{\rm max}}\frac{f\rmd f}{k\sub{h}(f)}=\nonumber\\
\Delta G_{X_{\rm min}}^{X_{\rm max}}-\frac{1}{2k_{\rm eff}}\left((f_{\rm max})^{2}-(f_{\rm min})^2\right) \,,
\label{g1mixeda}
\ee
we get
\be
\fl \Delta G_1(f_{\rm max})=\int_{0}^{x_{N}^{\rm max}}F_{\rm  ssDNA}^l(y)\rmd y-f_{\rm max}d_0+\Delta G_0\,, \qquad f_{\rm max}=F_{\rm  ssDNA}^{l_N}(x_{N}^{\rm max})\,.
\label{g1mixedb}
\ee
Note that the expression \eref{g1mixeda} for $\Delta G_1(f_{\rm max})$ is the equivalent of \eref{g1a} in the
force ensemble. According to \eref{g1mixedb} it only depends on the
maximum force at which the molecule is unfolded, therefore we
explicitely indicate such dependence of $\Delta G_1$ in
its argument. Using \eref{g1mixeda}, \eref{g1mixedb} it is possible to extract
$\Delta G_0$ from the knowledge of $\Delta G_{X_{\rm min}}^{X_{\rm
    max}},f_{\rm min} ,f_{\rm max}$ and the elastic properties of the
ssDNA:
\be
\fl \Delta G_{X_{\rm min}}^{X_{\rm max}}&=\frac{1}{2k_{\rm eff}}\left((f_{\rm
    max})^{2}-(f_{\rm min})^2\right)+\Delta G_1(f_{\rm max})
\label{wrevX3}\\
\fl \Delta G_0&=\Delta G_{X_{\rm min}}^{X_{\rm max}}-\frac{(f_{\rm
    max})^{2}-(f_{\rm min})^2}{2k_{\rm eff}}-G_{\rm ssDNA}(0\to x_{N};f_{\rm
    max})+f_{\rm max}d_0\,.
\label{wrevX4}
\ee

\end{document}